\documentclass[a4paper,12pt]{article}
\usepackage[utf8]{inputenc}
\usepackage[dvips]{color,graphicx}
\usepackage{mathrsfs}
\usepackage{epsfig}
\usepackage{amssymb}
\usepackage{amsmath}
\usepackage{feynmp}
\usepackage{slashed}
\usepackage{latexsym}
\usepackage{graphicx}
\usepackage{verbatim}

\newcommand{\be}{\begin{equation}}
\newcommand{\ee}{\end{equation}}
\newcommand{\ba}{\begin{array}}
\newcommand{\ea}{\end{array}}
\newcommand{\bea}{\begin{eqnarray}}
\newcommand{\eea}{\end{eqnarray}}
\newcommand{\balg}{\begin{align}}
\newcommand{\ealg}{\end{align}}
\newcommand{\bit}{\begin{itemize}}
\newcommand{\eit}{\end{itemize}}
\newcommand{\trm}[1]{\textrm{#1}}
\newcommand{\mbf}[1]{\mathbf{#1}}

\newcommand{\mcl}[1]{\mathcal{#1}}

\newcommand{\msc}[1]{\mathscr{#1}}

\newcommand{\nn}{\nonumber}

\begin{document}

\begin{titlepage}
\vspace*{-2cm}
\flushright{FTUAM 09-09\\ IFT-UAM/CSIC 09-27\\ULB-TH/09-15\\IFIC-09-22\\FTUV-09-0607}
\vskip 1.5cm
\begin{center}
{\Large\bf Minimal flavour seesaw models
\vspace{3mm}}
\end{center}
\vskip 0.5cm
\begin{center}
{\large M.B.~Gavela}$\,^a$~\footnote{belen.gavela@uam.es}, 
{\large T.~Hambye}$\,^b$~\footnote{thambye@ulb.ac.be},
{\large D.~Hern\'andez}$\,^a$~\footnote{d.hernandez@uam.es},
{\large P.~Hern\'andez}$\,^c$~\footnote{pilar@ific.uv.es}\\
\vskip .1cm
\vskip .7cm
$^a\,$ Departamento de F\'\i sica Te\'orica and Instituto de F\'\i sica Te\'orica UAM/CSIC,
\\
Universidad Aut\'onoma de Madrid, 28049 Cantoblanco, Madrid, Spain\\
\vskip .1cm
$^b\,$ Service de Physique Th\'eorique,\\
Universit\'e Libre de Bruxelles, 1050 Brussels, Belgium\\
\vskip .1cm
$^c\,$ Dpto. F\'\i sicaTe\'orica-UV and IFIC-CSIC, Edificio Institutos Investigaci\'on,\\
Apt 22805, E-46071 Valencia, Spain\\
\vskip .1cm
\end{center}
\vskip 0.5cm

\begin{abstract}
We explore realizations of  minimal flavour violation (MFV) for the lepton sector. We find that it can be realized 
within those seesaw models where a separation of the lepton number and lepton flavour violating scales can be 
achieved, such as  scalar mediated (type II) and inverse seesaw models. We present in particular a simple implementation of the MFV hypothesis which differs in nature from those previously discussed.
It allows to reconstruct the flavour structure of the model from the values of the light neutrino masses and mixing parameters, even in the presence of CP-violating phases. Experimentally reachable predictions for rare processes such as $\mu \rightarrow e \gamma$ are given.
\end{abstract}
\end{titlepage}
\setcounter{footnote}{0}
\vskip2truecm

\newpage
\section{Introduction}
Neutrino masses constitute the first evidence of physics beyond the Standard Model (SM). This 
new physics is likely to shed new light in the flavour puzzle, and could possibly be the  seed of the matter-antimatter asymmetry in the Universe. One of the most interesting questions is therefore whether this new physics can be tested through low-energy observables beyond neutrino oscillations, such as direct searches for the new particles involved,  rare decays or precision electroweak measurements. These effects  are however expected to be undetectable if the new physics scale  is orders of magnitude above the TeV, as is generally assumed.

In contrast, the possibility that the new physics scale is not too far beyond the electroweak scale opens new possibilities to test the origin of neutrino masses  in future experiments. In this context, however, the explanation of neutrino masses requires some symmetry principle to ensure their smallness as compared to the masses of other fermions. As it is well known, in the absence of new light degrees of freedom, the simplest symmetry principle that can ensure this suppression is global lepton number, which would forbid the Weinberg's effective operator responsible for light Majorana neutrino masses.  It is therefore conceivable that new dynamics exists that induces lepton flavour violation (LFV) at a scale $\Lambda_{FL}$, which could be as low as the TeV, while total lepton number (LN) is an approximate symmetry at this scale. The breaking of lepton number would result from subtler effects, which could be suppressed if they originate from a still higher energy scale $\Lambda_{LN}$ or if they are mediated by small couplings in a theory with only one scale. We will see that these two possibilities can be quite different  with respect to   
naturalness, but for the time being we will not distinguish between them: as long as those new scales are larger than the electroweak one, a model-independent representation is given in both cases by an effective theory of the type
\begin{eqnarray}
{\mathcal L} = {\mathcal L}_{SM} + {\alpha^{d=5} \over \Lambda_{LN}} {\mathcal O}^{d=5} + \sum_i {\alpha^{d=6}_i \over \Lambda^2_{FL}} {\mathcal O}_i^{d=6} + ... 
\label{eq:eft}
\end{eqnarray} 
In this expansion,  the only operator of dimension five ($d=5$) is Weinberg's \cite{eft}. The dimensionless couplings $\alpha^{d=5}, \alpha^{d=6}_i,...$ may be assumed to be of  ${\mathcal O}(1)$, while the effective scales $\Lambda_{FL}$, $\Lambda_{LN}$, take care of the suppressions of each type of contribution, with  $\Lambda_{LN} \gg \Lambda_{FL}$ as  required to obtain tiny neutrino masses. Therefore, all the effective couplings that break LN, such as Weinbergs's, are more suppressed than those that preserve lepton flavour symmetry, e.g.~those  of $d=6$. 

The phenomenology of the $d=6$ operators associated to neutrino masses has been extensively studied in the literature \cite{BGJ,unitarity,mfv1, mfv2,ABBGH}. Rare processes such as $\mu \rightarrow e \gamma$ can be generically quite large if the scale $\Lambda_{FL}$ is of $ \mathcal O$(TeV), but  
it is not possible to predict the strength of these processes in a model-independent way, since the flavour structure of the corresponding couplings, $\alpha^{d=6}_i$, is in general unrelated to that in neutrino masses, $\alpha^{d=5}$.

A class of models where a relation can be established are those incorporating {\it Minimal Flavour Violation} (MFV)~\cite{mfv1}. The Standard Model Lagrangian, ${\mathcal L}_{SM}$, would respect a large flavour symmetry group where it not for the presence of Yukawa couplings, $Y_u$, $Y_d$ (note that the presence of both types of coupling is necessary to induce physical flavour mixing in the quark sector)  and $Y_e$. MFV is the assumption that the only source of flavour violation in the full effective theory is the same as in ${\mathcal L}_{SM}$: ie. the Yukawa couplings, which  therefore 
should be included in the effective theory as flavour spurions.  

This hypothesis was first introduced in the context of the quark flavour sector \cite{mfvq}, and there it implies that the effective theory must be constructed  with the SM fields and the quark Yukawa couplings in order to satisfy the full flavour symmetry group of ${\mathcal L}_{SM}$. More precisely, the coefficients of the effective $d\ge6$ operators are specific combinations of the Yukawa couplings, which thus determine the flavour structure.  As a result, the theory avoids potentially too large flavour-violating effects and is very predictive  in the realm of flavour-violating processes. 

The same hypothesis in the lepton sector \cite{mfv1,mfv1bis} is more subtle, because strictly speaking the only breaking of lepton flavour in ${\mathcal L}_{SM}$ is due to the charged-lepton Yukawa couplings $Y_e$ that induce no flavour-changing effects by themselves (that is, in the absence of neutrino masses).  The additional flavour spurions needed to induce lepton mixing  are necessarily model dependent, since they must appear in the couplings of the lepton doublets to new fields. 
The authors of Ref.~\cite{mfv1} considered two such possibilities  in seesaw scenarios:
\begin{itemize}
\item Minimal case: the flavour spurions are the couplings of Weinberg's operator, i.e. $\alpha^{d=5}$ in eq.~(\ref{eq:eft}). As a consequence,  qualitatively speaking $\alpha^{d=6}\propto {\alpha^{d=5}}^\dagger \alpha^{d=5}$.
\item Extended case: there are very massive right-handed Majorana neutrinos (as in type I seesaw) and their Yukawa couplings to the lepton doublets provides the basic  flavour spurions. In the absence of CP-violating phases,  $\alpha^{d=6}\propto \alpha^{d=5}$.
\end{itemize}
Both assumptions imply interesting relations between the flavour structure of $d=5$ and $d=6$ couplings, or in other words between neutrino masses and rare processes such as $l_\alpha \rightarrow l_\beta \gamma$. The precise connection is different for the two cases.

The setup developed in Ref.~\cite{mfv1} assumes two fundamental -a priori unrelated- conditions to hold:
\begin{itemize}
\item[a)] Hierarchy between the operators that break and preserve lepton number or, in other words, a large hierarchy between the corresponding scales, $\Lambda_{FL} \ll \Lambda_{LN}$.
\item[b)] Flavour structure of the $d=6$ operator coefficients fixed by that of the $d=5$ one.
\end{itemize}

This setup rises however several fundamental questions.
In both extended and minimal MFV models, flavour spurions are introduced which are coupled to the physical fields responsible for the LN scale. How exactly can these spurions remain coupled, for example in the $d=6$ operator coefficients,  after the large scale $\Lambda_{LN}$ is integrated out? In order to fulfill conditions a) and b), is it necessary to have two distinct  scales, $\Lambda_{LN}$ and $\Lambda_{FL}$. Do these scales correspond to physical particle masses? Would this imply a naturalness problem \cite{natural,ABBGH}? How general are the relations found in Ref.~\cite{mfv1} between $d=5$ and $d=6$ operators in the extended case?

In this paper we address these questions by considering simple explicit  seesaw models that satisfy criteria a) and b).

Given that we consider  explicit models and not just some generic effective theory, we can distinguish two situations. Either condition b) is satisfied by the intrinsic structure of the model, or it  is a consequence of a restrictive MFV hypothesis.  Obviously the former case is more interesting and we will show  a couple of  examples of this type (in sections~\ref{sec:minimal} and \ref{sec:simplest}), 
where  the whole lepton flavour structure of the model can be extracted from the light neutrino mass matrix.  Furthermore, we will present a very simple model in section~\ref{sec:simplest} that satisfies conditions a) and b), but in which the relation between $d=5$ and $d=6$ operators is none of the kind considered in Ref.~\cite{mfv1}. For this model, no particular requirement about CP conservation is necessary.

\section{ MFV in scalar mediated (type-II) seesaw models}
\label{sec:minimal}

We are interested in explicit models fulfilling the two criteria a) and b) mentioned in the Introduction. In this section we stress that the type-II seesaw model is nothing but a MFV model of the minimal type (that is, where the basic flavour spurion is the coefficient of Weinberg's operator). It is the simplest example of such minimal MFV model.

As it is well known, the type-II seesaw model \cite{typeII} in its  basic form only adds to the SM fields one scalar hypercharge 2 scalar triplet field $(\delta^{++},\delta^+,\delta^0)$.  Writing this triplet as $(\,\frac{1}{\sqrt{2}}(\delta_1-i\delta_2),\delta_{3},\frac{1}{\sqrt{2}}(\delta_1+i\delta_2)\,)$, the most general Lagrangian can be easily written in terms of $\Delta\equiv(\delta_1, \delta_2, \delta_3)$:
\begin{eqnarray}
\!\!\mathcal{L}_\Delta&=& \left(D_{\mu}{\Delta}\right)^{\dagger}\left(D^{\mu} {\Delta}\right)+\left(\overline{\widetilde{\ell_\mathrm{L}}}Y_\Delta({\tau}\cdot{\Delta})\ell_\mathrm{L}
+{\mu_\Delta}{\widetilde{\phi}}^{\dagger}({\tau}\cdot{\Delta})^\dagger\phi
+ \text{h.c.}\right) -{\Delta}^{\dagger}{M_\Delta}^{2}{\Delta}\nonumber\\
\! \!\!&-& \frac{\lambda_{2}}{2}\left(\Delta^{\dagger}\Delta\right)^{2}-\lambda_{3}\left(\phi^{\dagger}\phi\right)\left(\Delta^{\dagger}\Delta\right)
-\frac{\lambda_{4}}{2}\left(\Delta^{\dagger}T^i\Delta\right)^{2}-\lambda_{5}\left(\Delta^{\dagger}T^i\Delta\right)\phi^{\dagger}\tau^i\phi
\label{ScalarL} \,,
\end{eqnarray}
with $\phi\equiv (\phi^+ \phi^0)^T$, $T_i$ being  the three-dimensional representation of the $SU(2)$ generators (as defined in Ref.~\cite{ABBGH}) and $\tau_i$ the Pauli matrices. In the absence of charged-lepton Yukawa couplings and $Y_\Delta$, the leptonic Lagrangian exhibits a global flavour symmetry group $SU(3)_L \otimes SU(3)_E $. 
The coexistence of $Y_\Delta$ and $\mu_\Delta$ explicitly  breaks lepton number, inducing at low energies the Weinberg operator:
\begin{equation}
\delta{\cal L}^{d=5} = \, c_{\alpha \beta}^{d=5} \,
\left( \overline{\ell_L^c}_{\alpha} \tilde \phi^* \right) \left(
\tilde \phi^\dagger \, {\ell_L}_{ \beta} \right) + {\rm h.c}.\, ,
\label{d=5}
\end{equation}
with 
\be\label{cd5scalar}
c^{d=5}_{\alpha \beta} = 2 Y_{\Delta\,\alpha \beta}\, \frac{ \mu_\Delta} {  M_\Delta^2} \,,
\ee  
which  yields a light neutrino Majorana mass matrix of the form ($v=246$~GeV) 
\be
m_{\nu\alpha \beta}=  -2Y_{\Delta\, \alpha \beta} \,\,v^2\,{\mu_\Delta \over { M_\Delta^2}} \,.
\label{mnutscal}
\ee
The coefficient of the $d=5$ operator is therefore proportional to $Y_\Delta$, which is the only flavour spurion of the model.  
As for the generated $d=6$ operators, there is only one at tree level which involves leptons \cite{ABBGH}\footnote{As shown in Ref.~\cite{ABBGH}, this model generates also two other $d=6$ operators involving scalar Higgs doublets and gauge bosons and no fermions, hence less interesting for our purpose since they do not carry any flavour structure.}:
\begin{equation}
\label{L4Ffamiliar}
\delta {{\cal L}^{d=6}}= c^{d=6}_{\alpha \beta \gamma \delta}
\left({\overline{\ell_L} _{\beta}}\gamma_\mu {\ell_L}_\delta \right)\left({\overline{\ell_L} _{\alpha}}\gamma_\mu {\ell_L}_{\gamma }\right)\,,
\end{equation}
with
\begin{equation}\label{cd6scalar}
c^{d=6}_{\alpha \beta \gamma \delta}=-\frac{1}{M_\Delta^2}{Y_\Delta}^{\dagger}_{\alpha \beta}
{Y_\Delta}_{\delta\gamma}\,. 
\end{equation}
 Note that the structure of $c^{d=6}$ is the generic one for $d=6$ leptonic operator coefficients in all seesaw models, $c^{d=6} \sim (M^{-1}Y)^\dagger M^{-1} Y$, where $Y$ and $M$ denote new Yukawas and scales, respectively~\cite{ABBGH}.
The comparison of eqs.~(\ref{cd5scalar}) and (\ref{cd6scalar}) shows that, in addition, the 
flavour structure of the  type II seesaw $d=6$  leptonic coupling goes basically like the square of that of the $d=5$ coupling, as in the minimal  MFV of Ref.~\cite{mfv1}.
In other words, in the type-II seesaw model if we know the flavour structure of the $d=5$ coefficient we know that of the $d=6$ ones. This is a well-known fact.   

In this framework, while the $d=5$ operator coefficient is proportional to $\mu_\Delta$, the $d=6$ coefficient is not. 
Therefore the decoupling in size of $d=5$ and $d=6$ couplings is automatic. With small enough $\mu_\Delta$, a tiny neutrino mass doesn't require large $M_\Delta$ and/or small Yukawa couplings $Y_\Delta$, hence the $d=6$ couplings can be sizeable. 
The only limit to this pattern is given by the rare decay constraints, as studied e.g.~in Ref.~\cite{ABBGH}. For example if $M_\Delta\sim 1$~TeV, $Y_\Delta\sim 10^{-1}$, $\mu_\Delta\sim 10^{-10}$~GeV, one gets neutrino masses of order $10^{-1}$~eV and saturates the experimental upper bound on 
 the $\mu \rightarrow e e e$ rate. The latter gives the most stringent constraint as $l\rightarrow 3l'$ decays are induced at tree level by the $d=6$ operator.

The phenomenological consequences of the relation $c^{d=6}_{\alpha \beta\gamma\delta} \propto c^{d=5\dagger}_{\alpha \beta} c^{d=5}_{\gamma \beta}$ have been studied in Ref.~\cite{Chun:2003ej}. Note that the fact that the operator in eq.~(\ref{L4Ffamiliar}) could emerge in the context of MFV theories has been raised in Ref.~\cite{mfv2}, independently of the type-II seesaw model.

The flavour breaking scale $\Lambda_{FL}$ is well defined in this case: it is the mass of the triplet. The lepton number violating scale $\Lambda_{LN}$ is instead  more subtly defined: 
 a large lepton number scale has been traded by a small $\mu_\Delta$ one, which does not correspond to the mass of any new physical particle. The $\Lambda_{LN}$ scale in eq.~(\ref{eq:eft}) would rather correspond now to the combination $\Lambda_{LN}\sim M_\Delta^2/\mu_\Delta$.
As the $\mu_\Delta$ term explicitly breaks lepton number (in conjunction with the dimensionless Yukawa coupling $Y_\Delta$), its small value is stable because $\mu_\Delta=0$ restores the lepton number symmetry. Therefore $\mu_\Delta$  does not necessarily require any large new physics scale to generate it. Alternatively, $\mu_\Delta$ could come from the spontaneous breaking of lepton number, i.e. from the vev $v_S$ of an extra scalar field. It could then be small owing to  a seesaw-type mechanism
i.e.~$\mu_\Delta \sim v_S^2/\Lambda'$ (in which case the scale of the new physics responsible for the small value of $\mu_\Delta$ could effectively be a large scale $\Lambda_{LN}=\Lambda'$), or  because 
$v_S$ is small and $\mu_\Delta=c\cdot v_S$ (with $c$ a dimensionless coefficient). 
Problems of stability of the scale $v_S$ are nevertheless to be expected in this framework with spontaneous breaking of lepton number, as discussed in Appendix A, unless the smallness of $\mu_\Delta$ is due to the smallness of the dimensionless coefficient $c$ rather than to the smallness of $v_S$.

In summary, the type-II seesaw model satisfies both criteria a) and b) above 
and to our knowledge there is no simpler model which satisfies them in a minimal--content minimal--flavour way.

\section{Two-scale fermionic mediated seesaw models (type-I and type-III)} 

In general all type I seesaw models are described by the following Lagrangian:
\begin{eqnarray}
{\cal L}&=& {\cal L}_{SM} + i \bar{N}_\alpha \!\not\partial N_\alpha- \left[ \lambda_N^{\alpha b} \bar{N}^\alpha \tilde{\phi}^\dagger \ell_L^b  +{M_{\alpha\beta} \over 2} \bar{N}^\alpha {N^{\beta c}}
+ h.c. \right] , 
\end{eqnarray}
giving rise to a neutrino mass matrix with the following block structure:
\begin{eqnarray}
 M_\nu = \left( \begin{array}{cc} 0 & \lambda_N^T v/\sqrt{2}\\  \lambda_N v/\sqrt{2} & M  
 \end{array} \right),  
\end{eqnarray}
where $\lambda_N$ is in general a $N \times 3$ matrix and $M$ is $N\times N$, with $N$ the number of sterile Weyl species. The lepton symmetry can be ensured for particular choices of the $\lambda_N$ and $M$ matrices.

In its minimal version~\cite{typeI}, there is only one new scale  encoded within the heavy right-handed neutrino mass matrix $M$,  
and since lepton number is violated by the simultaneous presence of $M$ and $\lambda_N$, we can identify it with $\Lambda_{LN}$. 
The flavour spurions, which in this case are the leptonic Yukawa couplings $\lambda_N$, 
would decouple when the heavy LN scale goes to infinity. It thus fails in satisfying condition a), which would require two distinct scales, and
it is not a valid model of MFV.

In order to achieve a successful MFV fermionic-mediated seesaw theory, some extra flavour dynamics at a lower scale, $\Lambda_{FL}$, is needed~\footnote{ For instance, this happens in type-I seesaw models with two scales built in. Recall as well that the scalar mediated type-II seesaw model in the previous section naturally encoded two distinct scales.}. Moreover, it is also necessary to identify the basic flavour spurions $\--$if there is more than one possible choice $\--$ and to  guarantee that in the limit $\Lambda_{LN}\rightarrow \infty$
 they remain coupled to the degrees of freedom active at the lower scale $\Lambda_{FL}$.

Type-I seesaw models with two scales built in do exist. 
It is well known  that type I seesaw models with suppressed $d=5$ and unsuppressed $d=6$  interactions  can be built, through the assumption of an approximately conserved lepton number  $U(1)_{LN}$ symmetry~\cite{u1,Branco:1988ex} (see  Refs.~\cite{pilaftsis,KS,ABBGH} for a recent discussion)\footnote{Seesaw models of type III \cite{typeIII} with unsuppressed $d=6$ operators can be constructed analogously \cite{ABBGH}. Since the phenomenology of flavour violating decays will be very similar, we restrict the explicit analysis to models with singlet fermions.}. 
The basic mechanism is to have a number of Weyl species such that those with opposite $U(1)_{LN}$ charges pair up into Dirac fermions,  while one or several charged species remain unpaired and therefore massless. The massless neutrinos only get masses when symmetry breaking interactions are included. The two scales are therefore related to the typical Dirac masses ($\Lambda_{FL}$) and 
the typical lepton number breaking scale ($\Lambda_{LN}$).  
At least two generic types of
flavour structures which do not decouple
in the limit of LN conservation, $\Lambda_{LN}\rightarrow \infty$, can be identified: 

\begin{itemize}
\item Type A: $\lambda_N$  and $M$  have the following block structures:
\begin{eqnarray}
\lambda_N^T = \left( \begin{array}{cc} Y_N^T & 0 \end{array}\right), \;\;\; M =   \left( \begin{array}{cc}0 & \Lambda^T  \\ \Lambda & 0   
\end{array} \right), 
\label{eq:typea}
\end{eqnarray}
In this case the $N=2 n$ sterile species divide in two groups with opposite lepton number charges, which we will denote by $N$ and $N'$.
The corresponding Lagrangian would read:
\begin{eqnarray}
{\cal L}_A&=& {\cal L}_{SM} + i \bar{N}^\alpha \!\not\partial N^\alpha+ i \bar{N'}^\alpha \!\not\partial {N'}^\alpha\nonumber\\
&-& \left[ Y_N^{\alpha b} \bar{N}^\alpha \tilde{\phi}^\dagger \ell_L^b  +{\Lambda_{\alpha\beta} \over 2} \left(\bar{N'}^\alpha {{N}^\beta}^c + \bar{N}^\beta {{N'}^\alpha}^c\right)
+ h.c. \right] . \nonumber\\
\label{eq:typea2}
\end{eqnarray}
Models of this type include those in Refs.~\cite{u1,Shaposh,blanchet}, often denominated {\it inverse} or {\it multiple}  seesaw models. The lepton number assignments are
$L_{N} = - L_{N'} = L_{\ell_L} = 1$. The pairs $(N^\alpha, N'^\alpha)$ combine into $n$  
massive Dirac fermions, while the 3 neutrinos remain massless for any $n$. 

\item Type B: $\lambda_N$ and $M$ have the following block structures:
\begin{eqnarray}
\lambda_N = \left( \begin{array}{ccc} Y^T_N & 0 & 0 \end{array}\right), \;\;\; M =   \left( \begin{array}{ccc}0 & \Lambda^T & 0  \\ \Lambda& 0 & 0 \\ 0 & 0 & \Lambda'     
\end{array} \right)\,, 
\label{eq:yukawab}
\end{eqnarray}
in which $M$ includes two distinct scales $\Lambda$ and $\Lambda'$ even in the lepton number conserving limit under discussion. 
The Lagrangian is then 
\begin{eqnarray}
{\cal L}_B & = & {\cal L}_{SM} + i \bar{N}^\alpha \slashed{\partial}N^\alpha+ i \bar{N'}^\alpha\slashed{\partial} N'^\alpha + i \bar{N''}^\alpha \slashed{\partial} {N''}^\alpha \nonumber\\
& - & \left[ Y_N^{\alpha b} \bar{N}^\alpha \tilde{\phi}^\dagger \ell_L^b  +{\Lambda_{\alpha\beta} \over 2} \left( \bar{N'}^\alpha N^{\beta c} +  \bar{N}^\beta N'^{\alpha c} \right) + {\Lambda'_{\alpha\beta} \over 2} \bar{N''}^\alpha N''^{\beta c} 
+ h.c. \right] , \nonumber\\
\label{eq:typeb}
\end{eqnarray}
and the lepton number assignments are $L_{N} = - L_{N'} = L_{\ell_L} = 1$ and $L_{N''} = 0$. 
 In this case therefore $N = 3n$, where $2n$ of the sterile species have opposite charges combining into $n$ massive Dirac fermions, as in model of Type A. The third group of $n$ massive Majorana singlets, $N''$, is decoupled again in the lepton number conserving limit, leaving behind 3 massless neutrinos.
It should be noted that the simplest example of type B model in eq.~(\ref{eq:yukawab}) corresponds to $n=1$. In this case,  $Y_N^T$ is a three-dimensional vector and $\Lambda$ and $\Lambda'$ are just numbers. This model has been recently discussed in Refs.~\cite{KS, ABBGH}, and it also corresponds to the structure
of the models considered earlier in Refs.~\cite{Branco:1988ex,pilaftsis}. 
\end{itemize}
Obviously there could be generalizations of the above to more species, but we  will discuss MFV in the context of these two possibilities.

The Lagrangian in eq.~(\ref{eq:typea2}) leads (for all $n$) to $n$ quasi Dirac fermions of masses $\sim \Lambda \gg v$ and 
three massless neutrinos that can get masses only if lepton number breaking entries are switched on.  Let us next consider how it can be implemented.


\section{The simplest  MFV  type-I seesaw model }
\label{sec:simplest}


We will now present the simplest possibility satisfying conditions a) and b), which will turn out to be  a model of type A with $n=1$.

Consider type A models above for general $n$. 
In order to obtain neutrino masses, it is necessary to break the $U(1)_{LN}$ symmetry, lifting the zeros in eq.~(\ref{eq:typea}). 
By naturalness arguments we should therefore lift all zeros at once. Let us then consider the matrix
\begin{eqnarray}
 M_\nu = \left( \begin{array}{ccc} 0 & Y^T_N v & \epsilon {Y}'^T_N v \\  Y_N v &  \mu' & \Lambda^T \\ \epsilon {Y}'_N v & \Lambda & \mu  
 \end{array} \right), 
 \label{eq:newperall}
\end{eqnarray}
where $\epsilon$ is a flavour-blind constant. 
$\epsilon, \mu$ and $\mu'$ are ``small parameters", that is,
the scales in $\mu, \mu'$ are much smaller than those in $\Lambda$ and $v$, and  $\epsilon \ll 1$,  to ensure an approximate $U(1)_{LN}$ symmetry. 
 
 The entry in the 22 element in eq.~(\ref{eq:newperall}) does not modify $c^{d=5}$ {\it at tree level}, and we will obviate it in what follows,  while entries in either the 13 or 33 elements do. 
 When the $n$ quasi Dirac neutrinos are integrated out, they give rise to both $d=5$ and $d=6$ effective operators (as expected in all type I seesaw models \cite{BGJ,ABBGH}):
\begin{eqnarray}
\delta {{\cal L}^{d=5}} & = &  c^{d=5}_{\alpha \beta}
\left({\overline{\ell_L^c} _{\alpha}}\tilde{\phi}^* \right)  \left(\tilde\phi^\dagger {\ell_L}_\beta\right)
\,, \label{Leff5typea1} \\
\delta {{\cal L}^{d=6}} & = & c^{d=6}_{\alpha \beta} ~\bar{\ell_L}^\alpha \tilde{\phi} i \slashed{\partial} \left( \tilde{\phi}^\dagger \ell_L^\beta \right) ,
\,
\label{Leff6typea2}
\end{eqnarray}
with coefficients~\footnote{  As recalled in the previous section,  the leptonic $c^{d=6}$ coefficients  are expected to depend on $(\Lambda^{-1} Y_N)^\dagger (\Lambda^{-1} Y_N)$, 
$(\Lambda^{-1} \epsilon {Y'}_N)^\dagger (\Lambda^{-1} Y_N)$ and $(\Lambda^{-1} \epsilon {Y'}_N)^\dagger (\Lambda^{-1} \epsilon {Y'}_N)$, and the last two contributions can thus be neglected at leading order in $\epsilon$.}
\begin{eqnarray}
\label{cstypea2}
c^{d=5}_{\alpha \beta} & \equiv & \epsilon \left({Y'_N}^T {1 \over \Lambda^T}Y_N + Y_N^T {1 \over \Lambda}{Y}'_N \right)_{\alpha\beta} - \left(Y_N^T {1 \over \Lambda} \mu {1 \over \Lambda^T}Y_N\right)_{\alpha\beta} \,, \\
c^{d=6}_{\alpha \beta}  & \equiv & \left(Y_N^\dagger {1 \over \Lambda^\dagger \Lambda}Y_N \right)_{\alpha\beta} \,+\, \mathcal{O}(\epsilon)\,.
\end{eqnarray}
Note that in general there is no relation between $c^{d=5}$ and $c^{d=6}$. However, we will see below that a direct connection does exists in the case $n=1$.  In this case,   $Y_N$ and $Y_N'$ are three dimensional complex column vectors, while $\Lambda, \mu$ and $\mu'$ are in general  complex numbers. This model  gives rise to just one massless neutrino, which is a viable possibility. 

In order to prove the connection between $c^{d=5}$ and $c^{d=6}$, we will start by showing that  in the case $\mu=\mu'=0$, we can reconstruct the Yukawa vectors $Y_N$ and $Y'_N$ (up to a global normalization) from $c^{d=5}$, and therefore we can fully predict the flavour structure of $c^{d=6}$. We will then show that the general case, eq.~(\ref{eq:newperall}) for $n=1$,  can be treated similarly.

Let us  then first consider the mass matrix
\begin{eqnarray}
 M_\nu = \left( \begin{array}{ccc} 0 & Y^T_N v & \epsilon Y'^T_N v \\  Y_N v &  0 & \Lambda^T \\ \epsilon Y'_N v & \Lambda & 0   
 \end{array} \right).  
 \label{eq:newper}
\end{eqnarray}
The $d=5$ and $d=6$ operator coefficients are then given by
\begin{eqnarray}
\label{cstypea}
c^{d=5}_{\alpha \beta} \equiv \epsilon \left({Y'_N}^T {1 \over \Lambda^T}Y_N + Y_N^T {1 \over \Lambda}Y'_N \right)_{\alpha\beta}, \,
c^{d=6}_{\alpha \beta} \equiv \left(Y_N^\dagger {1 \over \Lambda^\dagger \Lambda}Y_N \right)_{\alpha\beta}\,+\, \mathcal{O}(\epsilon)\, .
\end{eqnarray}
The texture in eq.~\eqref{eq:newper} has been considered previously in Ref.~\cite{Malinsky:2005bi}  for $n=3$. In that texture, lepton number is broken due to the simultaneous presence of all three types of terms, and light neutrino masses are then expected to depend on $Y_N$, $Y'_{N}$ and $\Lambda$. 
The flavour breaking in this model stems from both $Y_N$, $Y_N'$, and in consequence there is flavour violation even in the  lepton-number conserving $\epsilon \rightarrow 0$ limit, as $Y_N$ remains active in that limit:  non-trivial leptonic flavour physics can thus affect processes other than neutrino masses.

The structure of the effective Lagrangian in eq.~(\ref{eq:eft}) is therefore recovered if one identifies 
$\Lambda_{FL} \rightarrow \Lambda$ and $\Lambda_{LN} \rightarrow \Lambda/\sqrt{\epsilon}$.  The separation of scales is achieved by having a small $\epsilon$, which is technically natural since $\epsilon=0$ restores the lepton number symmetry.  The $\Lambda_{LN}$ scale does not correspond to any particle mass at this level, while $\Lambda_{FL}$ corresponds to the Dirac heavy right-handed neutrino mass scale, as expected. 

We will show that in this case the coefficient $c^{d=5}_{\alpha\beta}$ contains sufficient information to reconstruct {\it both} Yukawa vectors, up to a global normalization, and therefore also the flavour structure of $c^{d=6}_{\alpha\beta}$ up to a global normalization.  Furthermore,  this statement is valid even in the presence of CP violation, up to discrete degeneracies in the Majorana phases.  

It is easy to see how the number of real and imaginary parameters in the complete model actually matches those present in the effective operator coefficient $c^{d=5}$.  In particular, the number of physical phases in the light neutrino mass matrix is two, given that one neutrino is massless.
The fact that there are only two physical phases in the model is easy to see: the fundamental Lagrangian  has seven phases, three in $Y_N$, three in $Y'_N$ and one in $\Lambda$, and a rotation of the $N$, $N'$ fields and the three  lepton doublets gets rid of five of them. Only two physical phases remain and this also means that in the complete neutrino mass matrix there are only three yet unknown parameters: the angle $\theta_{13}$, the CKM type CP-violating phase ``$\delta$" and a unique Majorana phase ``$\alpha$". Furthermore, there is then a certain freedom  in the choice of basis for the complete theory, for instance it is possible to take real $\Lambda$ and $Y_N$  and also get rid of one of the 3 phases in $Y'_N$.
In what follows we will work in a basis in which $\Lambda$ is real while both $Y_N$ and $Y_N'$ may be taken as complex. 

Let us explicitly reconstruct the Yukawa couplings from the neutrino mass matrix. 
It is useful to introduce the notations:
\begin{eqnarray}
Y_N^T \equiv y \mbf{u}  \;\;\;  {Y'_N}^T \equiv y' \mbf{v}, \;\;\; 
\end{eqnarray}
where $y$ and $y'$ are real numbers and $\mbf{u}$ and $\mbf{v}$ are three complex vectors with unit norm. That is
\be
\langle \mbf{u} , \mbf{u} \rangle = \langle \mbf{v} , \mbf{v} \rangle = 1, 
\ee
where the scalar product is between complex vectors $\langle \mbf{u}, \mbf{v}\rangle \equiv \mbf{u}^\dagger \cdot \mbf{v}$.

The coefficient $c^{d=5}$ in eq.~(\ref{Leff5typea1}) can be rewritten as
\begin{eqnarray}
c^{d=5} &=& {\epsilon y y' \over \Lambda} \left(\mbf{u} \mbf{v}^T + \mbf{v} \mbf{u}^T \right) \equiv   {\epsilon y y'\over \Lambda}  \hat{O},\\
c^{d=6} &=& {y^2 \over \Lambda^2} \left( \mbf{u} \mbf{u}^{\dagger}  \right) + \mathcal{O}(\epsilon^2)\,. 
\end{eqnarray}
 Note that $c^{d=5}$ is symmetric in the exchange $\mbf{u}\leftrightarrow \mbf{v}$. This will result in discrete degeneracies of the Majorana phase $\alpha$, which cannot be resolved by the measurement of neutrino masses and mixing parameters.

$\hat{O}$ is a symmetric complex matrix and can therefore be diagonalized by a transformation of the form:
\begin{eqnarray}
{\epsilon y y' v^2 \over \Lambda}~U^T \hat{O} U = {\epsilon y y' v^2\over \Lambda}~\hat{O}_{d} \equiv - \left(\begin{array}{lll} m_1 & 0 & 0 \\
0 & m_2 & 0 \\
0 & 0 & m_3\end{array} \right),
\end{eqnarray}
where $m_i$ denote the mass eigenvalues, which are taken real, and $U$ is the unitary PMNS matrix. 

We can determine the mass eigenvalues and the entries of the $U$ matrix diagonalizing the hermitian matrix $\hat{O}^\dagger \hat{O}$, since
\begin{eqnarray}
U^\dagger \hat{O}^\dagger \hat{O} U = \hat{O}^2_{d} .
\end{eqnarray}
The three eigenvalues and eigenvectors of the matrix $\hat{O}^\dagger \hat{O}$ read:
\begin{align}
 \mu_0 = 0 \,,  &  \;\;\;\;    \mbf{e}_0 = \frac{\mbf{u} \times \mbf{v} }{ \sqrt{1 - |\mbf{u}\cdot\mbf{v}|^2}} \,, \;\;\; \\
 \mu_\pm = (1\pm \rho)^2  & \;\;\;  \mbf{e}_\pm = \frac{1}{\sqrt{2(1 \pm \rho)} } \left(e^{-i \theta/2}\mbf{u}^* \pm e^{i \theta/2}\mbf{v}^*\right) \,, 
\end{align}
where 
\be
\langle \mbf{u}, \mbf{v}\rangle =\langle \mbf{v}, \mbf{u}\rangle^* = \rho e^{i \theta}.  
\ee 

The PMNS matrix $U$ is now given by the matrix whose columns are precisely these eigenvectors \footnote{ Note that one mass is negative in our convention. That sign can be reabsorbed in a shift of the Majorana phase.}. Aside from discrete degeneracies in $\alpha$, the measurement of the neutrino masses and mixing parameters fully fixes then the eigenvectors and  allows to reconstruct the vectors $\mbf{u}$ and $\mbf{v}$ since:
\begin{eqnarray}
\mbf{u}^* = {e^{i\theta/2}\over \sqrt{2}} \left( \sqrt{1+\rho}~\mbf{e}_+ + \sqrt{1-\rho} ~\mbf{e}_-\right), \\
\mbf{v}^* = {e^{-i\theta/2}\over \sqrt{2}} \left( \sqrt{1+\rho} ~\mbf{e}_+ - \sqrt{1-\rho} ~\mbf{e}_-\right),
\end{eqnarray} 
while the ratio of the two mass splittings fixes $\rho$  (it quantitatively depends on the neutrino hierarchy).  The phase $\theta$ is not physical since it can be reabsorbed by 
rephasing the $N$ field by $e^{i\frac{\theta}{2}}$ and $N'$ by $e^{-i\frac{\theta}{2}}$ (leaving $\Lambda$ real)
and therefore we set it to zero for simplicity.

In order to do this matching precisely, we have to distinguish the cases of the two possible neutrino hierarchies.

\vspace{0.5cm}{\it Normal hierarchy}
\vspace{0.5cm}

In this case the ordering of the neutrino mass eigenstates is:
\be
m_1 = 0 \,, \quad |m_2|= {\epsilon y y' v^2\over \Lambda}~(1-\rho) \,, \quad  |m_3|= {\epsilon y y' v^2 \over \Lambda}~(1+\rho) \,, 
\ee
and therefore the columns of $U$ are ordered as $(\mbf{e_0},\mbf{e_-}, \mbf{e_+})$. 
From the ratio of the two neutrino splittings we can fix $\rho$:
\be
r \equiv {|\Delta m^2_{solar}|\over |\Delta m^2_{atmos}|} = {|\Delta m^2_{12}| \over |\Delta m^2_{23}|}
\;,\;\;\;\; \rho= \frac{\sqrt{1+r}-\sqrt{r}}{\sqrt{1+r} +\sqrt{r}} \,\,. 
\ee
Reading the columns of the PMNS matrix, one obtains 
\begin{eqnarray}
{Y_N}_i = { y\over \sqrt{2}} \left( \sqrt{1+\rho}~U^*_{i3} + \sqrt{1-\rho} ~U^*_{i2}\right)\,, \\
{Y'_N}_i = {y' \over \sqrt{2}} \left( \sqrt{1+\rho}~U^*_{i3} - \sqrt{1-\rho} ~U^*_{i2}\right). 
\end{eqnarray} 

We will use the standard angular parametrization of the PMNS matrix:
\begin{eqnarray}
U = \left(\begin{array}{lll}  c_{12} c_{13} & s_{12} c_{13} & s_{13} e^{-i\delta} \\
-s_{12} c_{23} - c_{12} s_{23} s_{13} e^{i\delta}  & c_{12} c_{23} - s_{12} s_{23} s_{13} e^{i \delta} & s_{23} c_{13}  \\
s_{12} s_{23} - c_{12} c_{23} s_{13} e^{i \delta} & -c_{12} s_{23} - s_{12} c_{23} s_{13} e^{i\delta} & c_{23} c_{13} \end{array}\right) U_{ph}
\label{eq:pmns1} 
\end{eqnarray}
where $U_{ph}$ contains the Majorana phases and can be parametrized in our case as:
\begin{eqnarray}
U_{ph} = \left(\begin{array}{lll}  e^{-i \alpha} & & \\
& e^{i \alpha} & \\
& & 1 \end{array}\right).
\end{eqnarray}
Up to terms of $\mcl{O}(\sqrt{r}, s_{13}) $, we find
\be
{Y^T_N} \simeq  y\left( \begin{array}{c}  e^{i\delta} s_{13} + e^{-i \alpha} s_{12}r^{1/4}  \\
s_{23} \left(1-{\sqrt{r}\over 2}\right) + e^{-i\alpha}r^{1/4}  c_{12}c_{23} \\ 
c_{23} \left(1-{\sqrt{r} \over 2} \right) - e^{-i\alpha}r^{1/4} c_{12}s_{23} \\ 
\end{array} \right) \,\,.  
\ee

Since the lightest neutrino is massless, from the central values of the atmospheric and solar parameters \cite{review},  we can also fix the combination
\begin{eqnarray}
\Big|{\epsilon y y' v^2\over \Lambda}\Big| \sim  0.029 \,\hbox{eV} \rightarrow \Big| {\epsilon y y'\over \Lambda} \Big| \sim 4.9 \times10^{-13} \,\hbox{TeV}^{-1} .
\end{eqnarray}

\vspace{0.5cm}
{\it Inverted hierarchy}
\vspace{0.5cm}

In this case the ordering of the neutrino mass eigenstates is:
\be
m_3 = 0\,, \quad  |m_1|=  {\epsilon y y'v^2\over \Lambda}~(1-\rho) \,, \quad  |m_2|=  {\epsilon y y'v^2\over \Lambda}~(1+\rho), 
\ee
and therefore the columns of $U$ are ordered as $(\mbf{e_-},\mbf{e_+}, \mbf{e_0})$. 
We find:
\bea
r = {|\Delta m^2_{12}| \over |\Delta m^2_{13}|} \; ,  \;\;\;\;  \rho= {\sqrt{1+r} - 1 \over \sqrt{1+r} + 1 } \;. 
\eea
and 
\begin{eqnarray}
{Y_N}_i = {y\over \sqrt{2}} \left( \sqrt{1+\rho}~U^*_{i2} + \sqrt{1-\rho} ~U^*_{i1}\right)\,, \\
{Y'_N}_i = {y'\over \sqrt{2}} \left( \sqrt{1+\rho}~U^*_{i2} - \sqrt{1-\rho} ~U^*_{i1}\right)\,.
\end{eqnarray} 
For the explicit  parametrization of the PMNS matrix $U$, we will use that in eq.~(\ref{eq:pmns1}).
Again, up to terms of $ \mcl{O}(\sqrt{r}, s_{13}) $ we find
\begin{eqnarray}
{Y^T_N} \simeq  {y\over \sqrt{2}} \left( \begin{array}{c} c_{12} e^{i\alpha}+ s_{12} e^{-i \alpha}  \\
c_{12} \left(c_{23} e^{-i\alpha} - s_{23} s_{13} e^{i(\alpha-\delta)} \right)  - s_{12} \left(c_{23} e^{i\alpha}+ s_{23} s_{13} e^{-i (\alpha+\delta)} \right) \\
-c_{12} \left(s_{23} e^{-i\alpha} + c_{23} s_{13} e^{i (\alpha-\delta)} \right)  + s_{12}  \left(s_{23} e^{i\alpha} - c_{23} s_{13} e^{-i (\alpha+\delta)}\right) \end{array} \right). 
\end{eqnarray} 

From the central values of the atmospheric and solar parameters \cite{review},  for the inverted hierarchy under study  it follows that 
\begin{eqnarray}
\Big| {\epsilon y y' v^2\over \Lambda} \Big| \sim  0.049 \, \hbox{eV} \rightarrow \Big| {\epsilon y y'\over \Lambda} \Big| \sim 8.1 \times10^{-13} \,\hbox{TeV}^{-1} .
\end{eqnarray}

\begin{figure}[tbp]
\begin{center}
 \includegraphics[width=7cm]{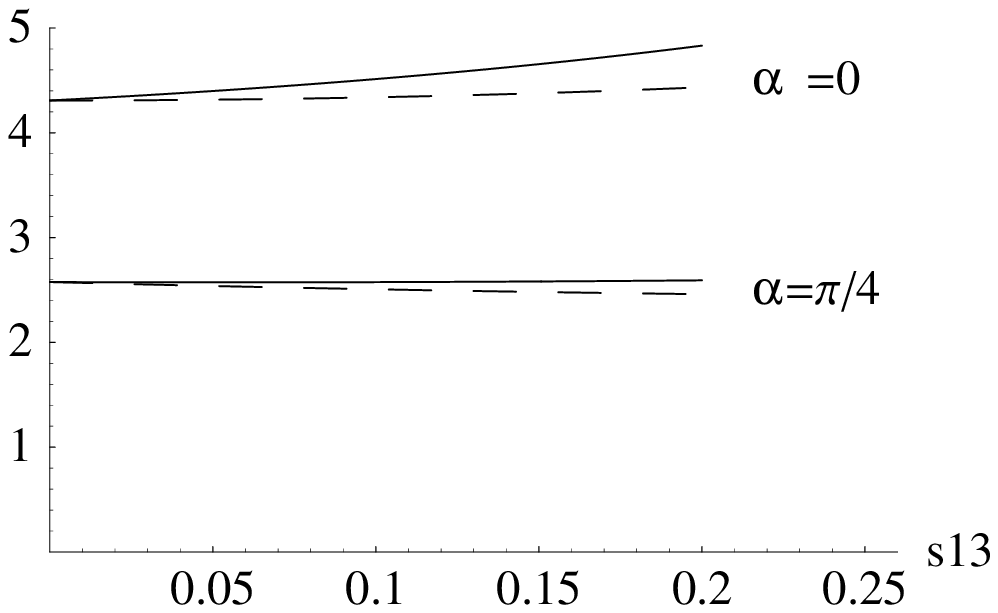} \includegraphics[width=7cm]{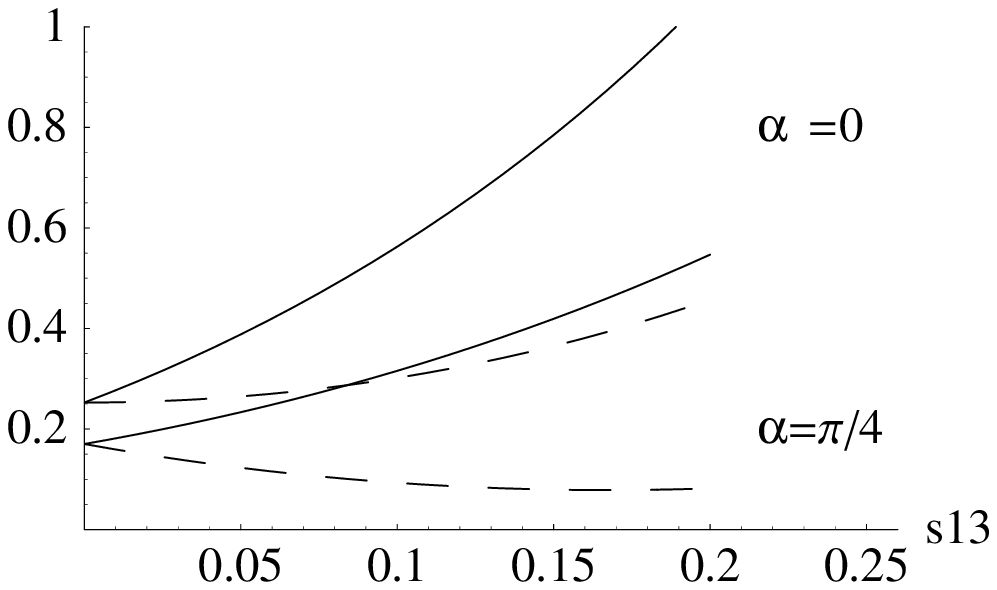}
\caption{ Normal hierarchy. Left: Ratio $B_{e\mu}/B_{e\tau}$  for different values of the CP phase $\delta=0$ (solid) and $\delta=\pi/2$ (dashed), with the two pairs of curves corresponding to $\alpha=0$ and $\alpha=\pi/4$ as 
denoted. Right: the same for the ratio  $B_{e\mu}/B_{\mu\tau}$ .} 
\label{fig:normal}
\end{center}
\end{figure}

\begin{figure}[tbp]
\begin{center}
\includegraphics[width=7cm]{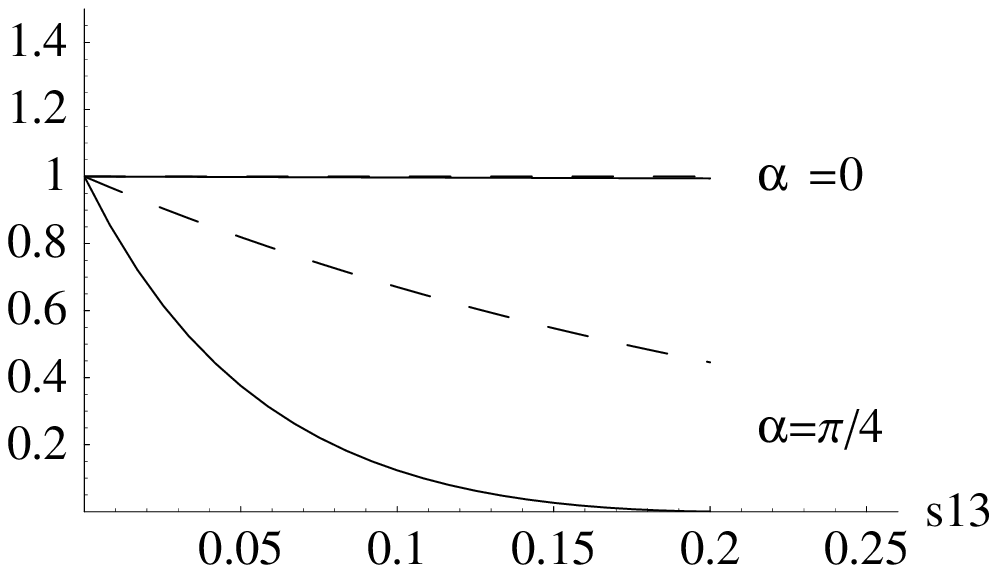} \includegraphics[width=7.5cm]{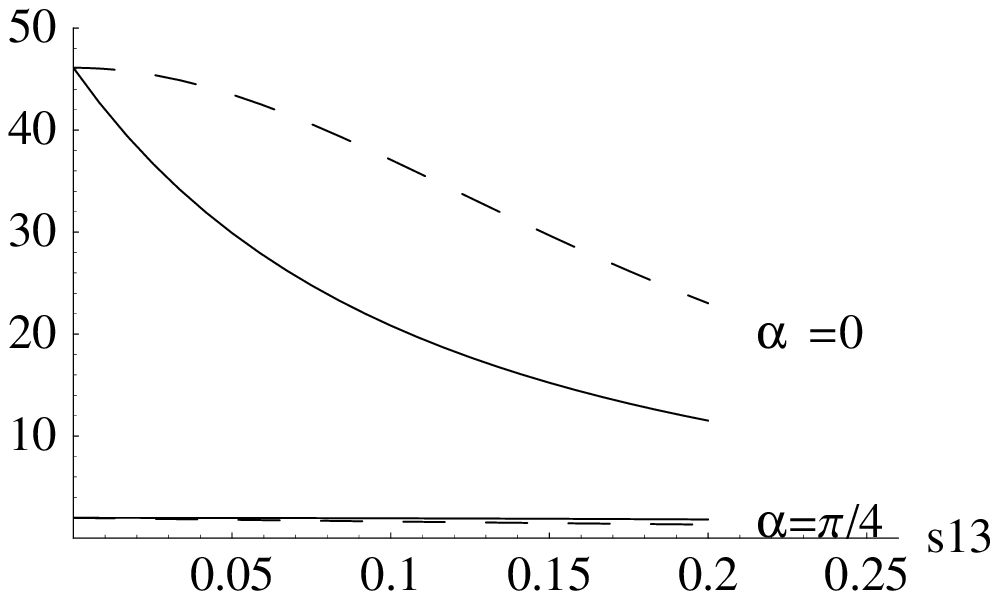}
\caption{ Inverted hierarchy. Left: Ratio $B_{e\mu}/B_{e\tau}$ for different values of the CP phase $\delta=0$ (solid) and $\delta=\pi/2$ (dashed), with the two pairs of curves corresponding to $\alpha=0$ and $\alpha=\pi/4$ as 
denoted. Right: the same for the ratio  $B_{e\mu}/B_{\mu\tau}$ .} 
\label{fig:inverse}
\end{center}
\end{figure}

Having reconstructed the full Yukawa vectors, it is now possible to make predictions for other lepton flavour violating processes.  It is interesting to estimate the rate for $l_i \rightarrow l_j \gamma$ processes and establish how do they depend on the unique free real parameter, $\theta_{13}$, and on the neutrino mass hierarchy. We will analyze the ratios 
\begin{eqnarray}
B_{ji} \equiv {\Gamma(l_i \rightarrow l_j \gamma) \over \Gamma(l_i \rightarrow l_j \nu_i \bar{\nu}_j) }\sim  |u_i^* u_j |^2
\,=\, \frac{1}{y^2}\, |Y_{N_i}Y_{N_j}|^2\,.
\end{eqnarray}
In Figs.~\ref{fig:normal} and ~\ref{fig:inverse} we show the results for the ratios $B_{e\mu}/B_{e\tau}$ and  $B_{e\mu}/B_{\mu\tau}$ as a function of $\theta_{13}$, for the normal and inverted hierarchies. 
The most striking feature is the strong dependence on the Majorana phase $\alpha$ of one of these ratios for both hierarchies: $B_{e\mu}/B_{e\tau}$ in the case of normal hierarchy, and $B_{e\mu}/B_{\mu\tau}$ for inverted hierarchy. 
In fact, within the ranges of $\delta$ and $\theta_{13}$ studied, the following prediction holds for the normal hierarchy:
\begin{eqnarray}
 B_{e\mu} \simeq \frac{9}{2} B_{e\tau}  &  \quad \alpha = 0 , \nonumber\\
 B_{e\mu} \simeq \frac{5}{2} B_{e\tau} &  \quad   \alpha = \pi/4,  \nonumber\\
  B_{e\mu} \simeq B_{e\tau}  & \quad  \alpha = \pi/2  \,.
\end{eqnarray}
while $B_{\mu\tau} > B_{e\mu}$. In contrast, a mild dependence on the $\delta$ phase holds for any $\theta_{13}$ value within the allowed range.

A different situation is found for the inverse hierarchy where, i.e. for vanishing $\theta_{13} = 0$, 
\begin{eqnarray}
B_{e\mu} \gg B_{\mu\tau}  &  \quad \alpha = 0 \,,  \nonumber\\
B_{e\mu} \simeq 2 B_{\mu\tau}  & \quad  \alpha = \pi/4 ,  \nonumber\\
B_{e\mu} \ll B_{\mu\tau}  & \quad\alpha = \pi/2 \,, 
\end{eqnarray} 
while $B_{e\mu} = B_{e\tau}$ holds. A significant dependence on $\delta$ may also develop for $\theta_{13} \ne 0$ for the two ratios considered depending on the value of the Majorana phase $\alpha$

\begin{figure}[htp]
\begin{center}
\includegraphics[width=7cm]{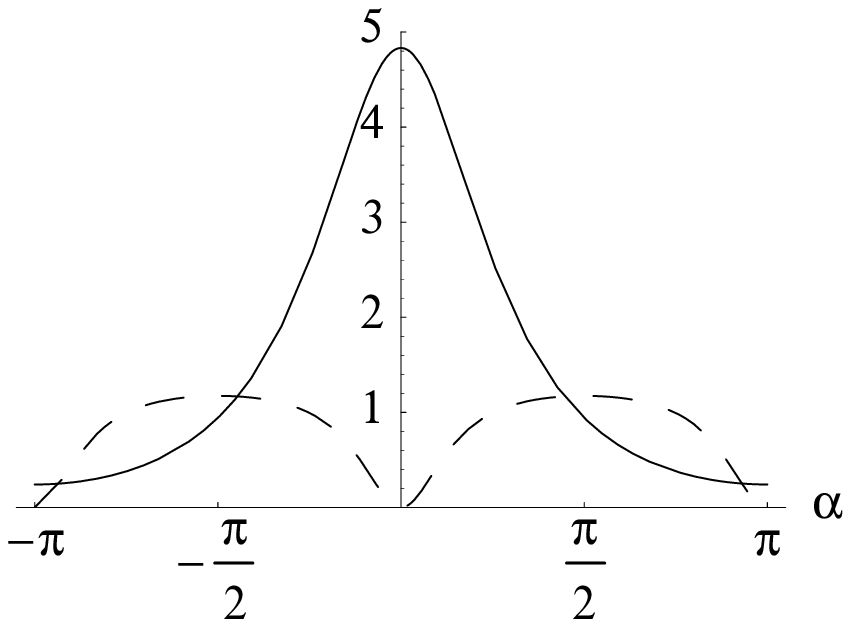} \includegraphics[width=7.5cm]{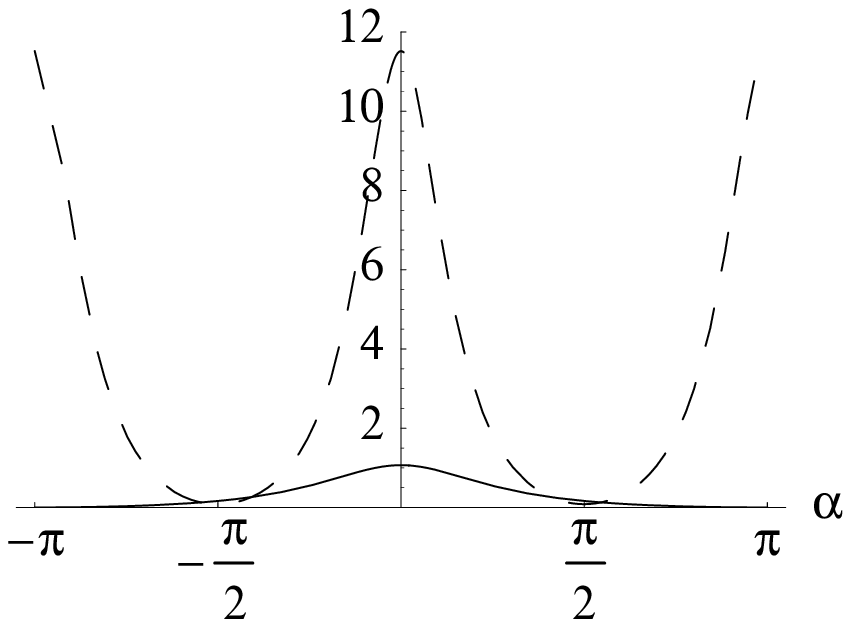}
\caption{ Left: Ratio $B_{e\mu}/B_{e\tau}$   for the normal hierarchy (solid) and the inverse hierarchy (dashed) as a function of $\alpha$ for $(\delta,s_{13})=(0,0.2)$.  Right: the same for the ratio $B_{e\mu}/B_{\mu\tau}$.} 
\label{fig:alpha}
\end{center}
\end{figure}

The $\alpha$-dependence of the ratios considered  has been plotted in Fig.~\ref{fig:alpha} for both hierarchies, for $\delta = 0, s_{13}=0.2$. 
\begin{figure}[tbp]
\begin{center}
\includegraphics[width=7cm]{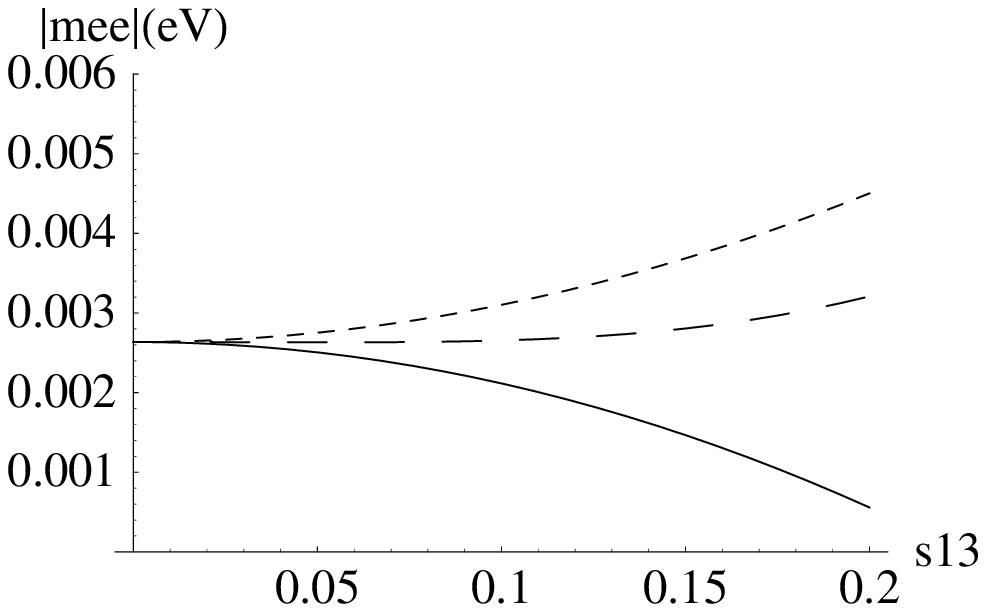} \includegraphics[width=7.5cm]{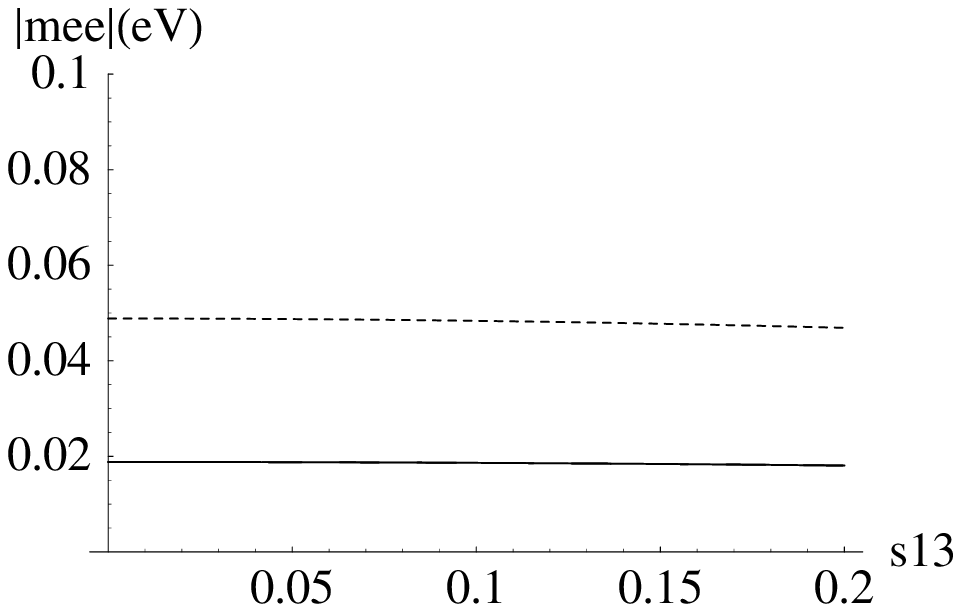}
\caption{ Left: $|m_{ee}|(eV)$ for the normal hierarchy as a function of $\sin \theta_{13}$ and for $(\delta,\alpha)=(0,0)$ (solid), $(0,\pi/4)$ (dotted) and $(\pi/2,0)$ (dashed). Right: the same for the inverse hierarchy.} 
\label{fig:bb0n}
\end{center}
\end{figure}
\begin{figure}[btp]
\begin{center}
\includegraphics[width=8cm]{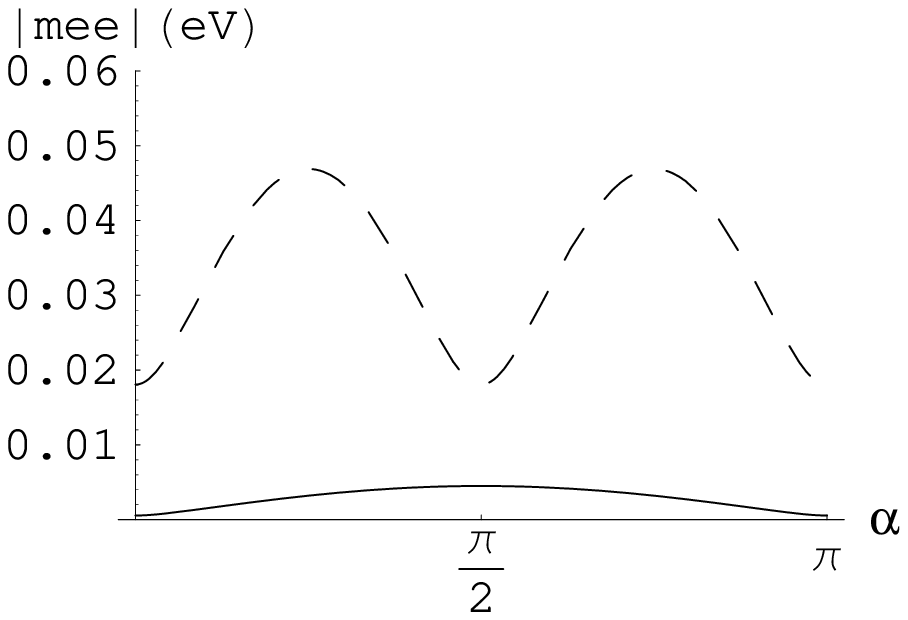}
\caption{$m_{ee}$ as a function of $\alpha$ for  the normal (solid) and inverted (dashed) hierarchies, for $(\delta,s_{13})=(0,0.2)$.} 
\label{fig:bb0nalpha}
\end{center}
\end{figure}

Note that the absolute normalization of the branching ratios is unconstrained, since neutrino masses only fix the combination $y y' v^2/\Lambda$, while the branching ratios depend on $y^2 v^2/\Lambda^2$. $\Lambda$ not far from the TeV scale is thus a viable possibility, and these branching ratios could therefore be measurable, provided $y'$ is small enough to account for the tiny neutrino masses.

\vspace{.2cm}

In Figs.~\ref{fig:bb0n} and \ref{fig:bb0nalpha} we show the expected value of $|m_{ee}|$ to be measured in neutrinoless  beta decay, for the normal and inverse hierarchies and for the central experimental values of the known parameters as a function of $s_{13}$ and $\alpha$. Note that these figures show degeneracies in the value of $\alpha$ that can be resolved from the measurement of the $c^{d=6}$ couplings, i.e. from the radiative decays discussed above. As expected, the value of $|m_{ee}|$ is of ${\mathcal O}(10^{-3} eV)$ for the normal hierarchy and one order of magnitude above for the inverse one. Expanding in the small parameters $s_{13}$ and
$r^{1/2}$, the following approximate expressions result (taking the central values for $s_{23}$ and $s_{23}\simeq c_{23}$):
\begin{align}
\left|m_{ee}\right|_{NH} & \simeq 0.058 ~eV~ \left| s_{13}^2e^{2i\delta}- s_{12}^2 e^{-2i\alpha}\sqrt{r}(1-\sqrt{r}) \right| \nonumber \\ 
\left|m_{ee}\right|_{IH} & \simeq   0.049 ~eV~ \left| s_{12}^2e^{-2i\alpha} - c_{12}^2 e^{ 2 i \alpha}  \right| + \mathcal{O}(r, s_{13}^2) \;. 
\end{align}
The inverse hierarchy case is in consequence approximately independent of $s_{13}$ and therefore of the CKM-like phase $\delta$, but very sensitive to the Majorana phase $\alpha$. In the normal hierarchy case, the dependence on all the parameters is significant. In both cases, it is important to stress that the measurement of $|m_{ee}|$, together with that of the
neutrino mixing parameters in future neutrino oscillation experiments can in principle fix {\it all} the parameters of the  model, except the absolute normalization of the $d=6$ operator~\footnote{ 
 Note also that the relation between the $d=6$ and $d=5$ flavour structures obtained above is not of the ``minimal" or ``extended" MFV types and is not based on the assumption of an underlying flavour symmetry (such as a $O(n)$ symmetry enforced in Ref.~\cite{mfv1}  to have a right-handed neutrino mass matrix proportional to the identity).}.

\vspace{0.5cm}

Let us now turn to the more general case when $\mu, \mu' \neq 0$ in eq.~(\ref{eq:newperall}).  It turns out that {\it all} the results previously derived in this section hold as well for this general case. This can be easily seen by noting that, for the corresponding $c^{d=5}$ coefficient in eq.~\eqref{cstypea2}, 
\begin{eqnarray}
c^{d=5}_{\alpha \beta}& = & \epsilon \left(Y_N'^T {1 \over \Lambda^T}Y_N + Y_N^T {1 \over \Lambda}Y'_N \right)_{\alpha\beta} - \left(Y_N^T {1 \over \Lambda} \mu {1 \over \Lambda^T}Y_N\right)_{\alpha\beta} \nonumber\\
&=&  \epsilon \left[\left(Y'_N - {k\over 2} Y_N\right)^T {1 \over \Lambda^T}Y_N + Y_N^T {1 \over \Lambda} \left(Y'_N - {k\over 2}Y_N \right)\right]_{\alpha\beta} ,  \, 
\end{eqnarray}
with 
\begin{eqnarray}
k \equiv {1 \over \epsilon} \mu {1 \over \Lambda^T}. 
\end{eqnarray}
Therefore $c^{d=5}$ has the same structure of that in eq.~(\ref{cstypea}) with the substitution
\begin{eqnarray}
Y'_N  \longrightarrow Y'_N - {k\over 2} Y_N . \;\;\; 
\label{eq:ytilde}
\end{eqnarray} 
We can consequently reconstruct $Y_N$ and the combination in eq.~(\ref{eq:ytilde}) from the neutrino mass matrix,  that is from $c^{d=5}$, exactly as we did before. 
  From these two combinations, we cannot reconstruct $Y'_N$ in eq.~\eqref{eq:newperall} ,  because the factor $k$ is a new free parameter. Nevertheless, all the flavour violating processes induced by $c^{d=6}$ depend only on $Y_N$, at leading order in the lepton-number violation parameters, and are therefore the same. In other words, the structure in eq.~\eqref{eq:newperall}  is as predictive as that in eq.~\eqref{eq:newper}. 
The low-energy physics (i.e. the relation between flavour violation transitions and the neutrino mass matrix) is the same in both models.

\vspace{0.5cm}

A nice feature of the model considered in this section, eq.~(\ref{eq:newperall}), is its naturalness characteristics. It does not contribute significantly to the electroweak hierarchy problem for $\Lambda$ values near the $TeV$ scale, as all loop corrections relevant to Higgs physics are proportional to small parameters.

Finally, given the predictivity of the model, it would be interesting to explore whether it leads to successful leptogenesis. At low scale, a small mass splitting between the right-handed neutrinos is necessary in order to have a large resonant enhancement of the CP-asymmetry. This indeed happens in the model discussed here, eq.~(\ref{eq:newperall}), which induces a tiny mass difference of order of the size of the $U(1)_{LN}$ breaking, and hence leads to a large resonant enhancement (with however e.g.~large washout effects from inverse decays and $\Delta L=2$ scatterings for large values of the $Y_N$ couplings).
This has been analyzed in ref.~\cite{blanchet} for the case where the 22 and 33 entries in eq.~(\ref{eq:newperall}) dominate the mass splitting. Successful leptogenesis appears to be achievable in this case, although only for relatively small values of all Yukawa couplings, which in turn leads to  suppressed flavour changing $d=6$ effects, even for a mass splitting at the resonance peak.
The case with negligible 22 and 33 entries, eq.~(\ref{eq:newper}),
is yet to be analyzed.


\vspace{0.5cm}

We will consider next an alternative class of candidate MFV models: those in which  lepton number violation results from lifting the zeros in the diagonal entries of the $M_\nu$ matrix, with no 13 entry  and $n>1$. These are the well known inverse seesaw models \cite{u1}.

\section{MFV in type-I inverse seesaw models}
 
This section deals, as did the previous one, with models of type A, see eq.~(\ref{eq:typea}). We consider now the case in which light neutrino masses result from lifting the zeros in the diagonal entries of $M_\nu$. In contrast to the case with only off-diagonal lepton-violating entries, eq.~\eqref{eq:newper}, the diagonal entries are soft-breaking terms and therefore 
 would not induce by themselves off-diagonal terms.  
The fundamental neutrino mass matrix is of the form:
\begin{eqnarray}
 M_\nu = \left( \begin{array}{ccc} 0 & Y^T_N v/\sqrt{2} & 0 \\  Y_N v/\sqrt{2} &  \mu' & \Lambda^T \\ 0 & \Lambda & \mu   
 \end{array} \right)\,.
 \label{MnuA} 
\end{eqnarray}
 For $n=1$ however it leads to two massless neutrinos and in consequence is of no physical interest. $n\geq 2$ is needed to get at least 2 massive neutrinos~\cite{u1}. 
 The simultaneous  presence of $Y_N$, $\Lambda$ and the Majorana couplings $\mu$ and/or $\mu'$ breaks lepton number. 
 As explained before the $\mu'$ scale does not play any role at low-energies at tree level. 

 The tree-level exchange of the heavy species gives rise to the same  $d=5$ and $d=6$ effective operators in 
eqs.~(\ref{Leff5typea1})-(\ref{Leff6typea2})  with coefficients
\begin{eqnarray}
\label{cstypeis}
c^{d=5}_{\alpha \beta} \equiv - \left(Y_N^T {1 \over \Lambda} \mu {1 \over \Lambda^T}Y_N\right)_{\alpha\beta} , \;\;\;\;c^{d=6}_{\alpha \beta} \equiv \left(Y_N^\dagger {1 \over \Lambda^\dagger \Lambda}Y_N \right)_{\alpha\beta} .
\end{eqnarray}
The structure of the effective Lagrangian in eq.~(\ref{eq:eft}) is therefore recovered if one identifies 
$\Lambda_{FL} \rightarrow \Lambda$ and $\Lambda_{LN} \rightarrow \Lambda^2/\mu$.  The separation of scales is achieved by having a small $\mu$, which is technically natural since $\mu=0$ restores the lepton number symmetry.  

Concerning the flavour structure of the $d=5$ and $d=6$ operators in eq.~(\ref{cstypeis}),  they are, in general, unrelated. That is, unless $\mu \sim I_{n\times n}$,  which amounts to saying that the term preserves an additional $O(n)$ symmetry.  Obviously this symmetry is broken by the $Y_N$ and $ \Lambda$ couplings, and in consequence it can be argued that  there is a priori no justification for this choice, which will not be stable under radiative corrections.  Nevertheless, this choice is equivalent to the assumption or hypothesis of MFV:  that the only sources of flavour violation are encoded
in the charged lepton Yukawa coupling, $Y_e$, 
  in $Y_N$ and maybe also in $\Lambda$.  If these three couplings were zero, then  the lepton sector would  have a symmetry group: 
\begin{eqnarray}
SU(3)_{\ell_L} \times SU(3)_E \times SU(n)_{N} \times O(n)_{N'} .
\label{symA1}
 \end{eqnarray}
Alternatively, the option  $\lambda_E=Y_N=0$ with $\Lambda$  proportional to the identity would imply that the flavour symmetry group is
 \begin{eqnarray}
SU(3)_{\ell_L} \times SU(3)_E \times O(n)_{N,N'} .
\label{symA2}
 \end{eqnarray}
In the former case the neutrino sector spurions are $Y_N \sim (\bar{3},1, n, 1)$ and  $\Lambda \sim (1, 1, n, n)$ , while in the latter $Y_N \sim (\bar{3},1,n)$. In both cases, 
 the exact connection of $d=5$ and $d=6$ couplings {\it only holds  up to 
 CP phases}. Indeed, in the absence of CP violation it follows that 
 \begin{eqnarray}
 c^{d=5}_{\alpha \beta} = - \mu ~c^{d=6}_{\alpha\beta},
 \end{eqnarray} 
and the flavour processes induced by the $d=6$ operator are fixed, up to a global normalization, by the neutrino mass matrix.  This model with diagonal $\mu$ is therefore the simplest example of the extended class of models defined in Ref.~\cite{mfv1}.

In Refs.~\cite{mfv1,mfv2}, the implications for  flavour-violating processes $l_i \rightarrow l_j \gamma$ as well as $\mu e$ conversion in extended models of MFV have been discussed and should apply as well to the model discussed here. However, it turns out that the $d=6$ Lagrangian at tree level contains just one operator, eq.~(\ref{Leff6typea2}), which is none of those appearing in  the basis considered in Ref.~\cite{mfv1}. It can obviously be rewritten in terms of operators in that list:
\begin{eqnarray}
 \delta {{\cal L}^{d=6}}=c^{d=6}_{\alpha \beta} ~\bar{\ell_L}^\alpha \tilde{\phi} i \slashed{\partial} \left( \tilde{\phi}^\dagger \ell_L^\beta \right) = {c^{d=6}_{\alpha \beta} \over 2} \left(\bar{\ell_L}^\alpha \gamma_\mu \ell_L^\beta ~\phi^\dagger i D_\mu \phi - \bar{\ell}_L^\alpha \mbf{\tau} \gamma_\mu \ell_L^\beta ~ \phi^\dagger \mbf{\tau} i D_\mu \phi \right).  
 \label{eq:blind}
\end{eqnarray}
 The combination is however a {\it blind} direction:   $l_i \rightarrow l_j \gamma$ and $\mu \rightarrow e$ do not take place at tree level, as it happens separately for any of the two operators on the right -hand side of eq.~(\ref{eq:blind}), but only at one loop. In consequence, the bounds derived  from these processes in Refs.~\cite{mfv1,mfv2} are further suppressed by an additional loop factor, roughly $1/(4 \pi)^2 \sim 10^{-2}$. The flavour structure is however the same. Similar plots to those shown in Figs.~\ref{fig:normal}, ~\ref{fig:inverse} can be found in Ref.~\cite{mfv1}, which should be strictly applicable to our case. They found the pattern $B_{\mu\tau} \gg B_{e\mu} \sim B_{e\tau}$, which is to be contrasted with the findings in the previous section. 

\vspace{0.5cm}

Also in this case it is necessary to justify the presence of the  $\mu, \mu'$ terms and no other $U(1)_{LN}$ breaking term, such as for instance a $13$ entry in eq.~(\ref{MnuA}) as in the model in previous section.  The symmetry pattern shown in eq.~(\ref{symA1}) could justify it. Alternatively, such a choice  could be justified if  the $U(1)_{LN}$ symmetry is spontaneously broken by the vacuum expectation value (vev) of a scalar singlet $S$ with charge -2, leading to a Lagrangian of the form:
  \begin{eqnarray}
{\cal L}_A&=& {\cal L}_{SM} + i \bar{N} \!\not\partial N + i \bar{N'} \!\not\partial {N'}
- \left[ Y_N \bar{N} \tilde{\phi}^\dagger \ell_L  +{\Lambda \over 2} \left(\bar{N'} {{N}}^c + \bar{N} {{N'}}^c\right) \right. \nonumber\\
&+& \left.{ g S \over 2}  \bar{N'} N'^c  + { g' S^\dagger \over 2}  \bar{N} N^c+h.c. \right]  + V(S, \phi). 
\label{eq:typea_singlet}
\end{eqnarray}
A vev of the singlet  would induce the $\mu$ and $\mu'$ couplings $\mu = g \langle S \rangle, \mu' = g' \langle S^\dagger \rangle$. 
 Nevertheless, this possibility results in a naturalness problem, that is, of the stability of the separation of scales at the quantum level. We discuss it briefly in appendix A. 


\section{MFV in type-I seesaw models of type B}

The models of type B, e.g.~with $3n$ sterile species, also satisfy an exact global $U(1)_{LN}$ symmetry, which ensures the presence of three massless neutrinos for any value of $n$.  In order to lift their masses it is necessary to have some entries in the mass matrix that violate the symmetry. 
There are several possibilities with different implications  in what respects MFV. One possibility is to include some small entries in the zeros of  $M$. The modification of only the diagonal entries in $M$ reduces the model to one of type A, since the 
$N''$ fields would remain decoupled in this case. The modification instead of only the off-diagonal entries induces a neutrino mass matrix of the form:
\begin{eqnarray}
 M_\nu = \left( \begin{array}{cccc} 0 & Y_N v/\sqrt{2} & 0 & 0 \\  Y_N^T v/\sqrt{2} &  0 & \Lambda  & \mu_2 \\
  0 & \Lambda^T  & 0 & \mu_1 \\
  0 & \mu_2 & \mu_1 & \Lambda' \\
    \end{array} \right). 
    \label{mnutypeb}
\end{eqnarray}
The main interest of these models, 
in comparison with models of type A, is that it is no longer necessary to assume that $\mu_1$ and $\mu_2$ are very small scales.  Even more, in the limit in which $\Lambda' $ is much larger than all the other scales present, it reduces to a Type A model. In other words, type B models can be seen as an ultraviolet completion of type A scenarios, whose small scales are then explained in terms of large ones in the  fundamental theory. Let us discuss this point in detail.

The separation of scales, that is, the implementation of criterium a) in the Introduction, 
can be achieved through a hierarchy of scales:
$\Lambda ' \gg \Lambda, \mu_1, \mu_2$. In principle $\mu_1$ and $\mu_2$ could be roughly  $\sim \Lambda$, because the $U(1)_{LN}$ symmetry is recovered when the scale $\Lambda'$ decouples, no matter how large are the other scales. Indeed, integrating out the scale $\Lambda'$, 
the effective theory at energies below $\Lambda'$ is:
\begin{eqnarray}
{\mathcal L}_B&\simeq& {\cal L}_{SM} + i \bar{N} \!\not\partial N + i \bar{N'} \slashed{\partial} {N'} 
-  \left[ Y_N\bar{N} \tilde{\phi}^\dagger \ell_L  +  {1 \over 2} \left( \Lambda + \mu_2 {1\over \Lambda'} \mu_1^T \right) \left( \bar{N'} N^c +  \bar{N} {{N'}}^c \right) \right. \nonumber\\
&+&\left. {1\over 2} \mu_2 {1\over \Lambda'} \mu_2^T  \bar{N} {{N}}^c +{1\over 2} \mu_1 {1 \over \Lambda'} \mu_1^T\bar{N'} {{N'}}^c
+ h.c. \right] .
\end{eqnarray}
This is nothing but a  model of type A,  with symmetry-breaking entries of the $\mu$, $\mu'$ type, in eq.~(\ref{MnuA}) suppressed by the large scale $\Lambda'$.  
The scale of lepton number violation can be simply identified with $\Lambda_{LN} \sim \Lambda'$, which corresponds to the mass of the heavy Majorana neutrinos, while the scale of lepton flavour violation would be $\Lambda_{FL}\sim \Lambda$. This pattern is close to that of the extended models of Ref.~\cite{mfv1}.

When the scale $\Lambda$ is sufficiently above the electroweak scale, it can be integrated out, resulting in the same  d=5 and  d=6 operators than in eq.~(\ref{cstypeis}), with $\mu$ given now by ${\mu_1 {1\over \Lambda'} \mu_1^T}$. The effective theory at scales much lower than $\Lambda$ is therefore:
\begin{eqnarray}
{\mathcal L}_B&\simeq& {\cal L}_{SM}  - \left(Y_N^T {1 \over \Lambda} {\mu_1 {1\over \Lambda'} \mu_1^T}  {1 \over \Lambda^T}Y_N\right)_{\alpha\beta} \left({\overline{\ell_L^c} _{\alpha}}\tilde{\phi}^* \right)  \left(\tilde\phi^\dagger {\ell_L}_\beta\right) \nonumber\\
& + & \left(Y_N^\dagger {1 \over \Lambda^\dagger} {1 \over \Lambda}Y_N\right)_{\alpha\beta}  ~\bar{\ell_L}^\alpha \tilde{\phi} i \not\!\partial \left( \tilde{\phi}^\dagger \ell_L^\beta \right) + \mathcal{O}\left({1 \over \Lambda'^2}, {1\over \Lambda^2 \Lambda'} \right)\, ,
\label{LtypeB}
\end{eqnarray}
to be compared with the typical structure of inverse seesaw models, eq.~(\ref{cstypeis}). 
The $c^{d=6}\propto c^{d=5}$ relation between the flavour structures of d=5 and d=6 operators discussed in section 4 holds (up to CP phases),  provided we assume that the flavour symmetry group is
\begin{eqnarray}
SU(3)_{\ell_L} \times SU(3)_E \times SU(n)_{N} \times O(n)_{N',N''} , 
\label{typeBsym1}
\end{eqnarray}
and is only broken by the spurions $Y_N \sim (\bar{3}, 1, n, 1)$ and $\Lambda \sim (1, 1, n, n)$, while both $\Lambda'$ and $\mu_1$ are invariant under $O(n)$ rotations of the $N''$ and $N'$ fields. In this situation, $\mu_2 \sim (1,1,n, n) \sim \Lambda$.  Would $\Lambda$ be instead proportional to the identity and $Y_N$ the only spurion, then the symmetry group would be 
\begin{eqnarray}
SU(3)_{\ell_L} \times SU(3)_E \times O(n)_{N,N',N''} , 
\label{typeBsym2}
\end{eqnarray}
and $\mu_2$ would also be proportional to the identity. 

Concerning the justification of the zeros in eq.~(\ref{mnutypeb}), we 
note that the flavour symmetries  just described are not enough to forbid, 
for example, a 33 entry in the case of eq.~(\ref{typeBsym1}), or  13 and 
14 entries (proportional to $Y_N$) in the case of eq.~(\ref{typeBsym2}). 
However , it is easy to justify a breaking of the $U(1)_{LN}$ symmetry 
only through the $\mu_1$ and $\mu_2$ terms, if we assume that the symmetry 
has been spontaneously broken through the vev of a singlet scalar $S$ with 
lepton number $L_S=+1$. The only possible renormalizable couplings of the 
scalar  to fermions would then be precisely those giving rise to the 
$\mu_1$ and $\mu_2$ terms, see eq.~(\ref{LwithScaseB}) in Appendix~A. As 
in the type A models with spontaneous symmetry breaking, questions of 
naturalness may arise though, as we briefly discuss in that appendix.


As in the case of type A models, an alternative to break the global symmetry is to lift the zeros in $\lambda_N$, that is  the 13 or 14 entries in the neutrino matrix in eq.~(\ref{mnutypeb}). A  $13$ entry  would reduce the model at low energies to that discussed in section~3. On the contrary, a  $14$ entry would be qualitatively different:
\begin{eqnarray}
 M_\nu = \left( \begin{array}{cccc} 0 & Y_N v/\sqrt{2} & 0 & Y_N' v/\sqrt{2} \\  Y_N^T v/\sqrt{2} &  0 & \Lambda  & 0 \\
  0 & \Lambda^T  & 0 & 0 \\
  Y_N'^T v/\sqrt{2} & 0& 0 & \Lambda' \\
    \end{array} \right), 
\label{MtypeB2}     
\end{eqnarray}
with $Y_N'$ and $Y_N$ being distinct spurions, since the quantum numbers of $N_\alpha$ and $N''_\alpha$ are different. The approximate $U(1)_{LN}$ symmetry is ensured in this case not by a suppressed $Y'_N$, but rather by a large
hierarchy $\Lambda' \gg \Lambda$. The integration of the scale $\Lambda'$ and $\Lambda$ in this case gives now rise to  the d=5 and d=6 operators with coefficient matrices given by:
\begin{eqnarray}
\label{cstypeb}
c^{d=5}_{\alpha \beta} \equiv \epsilon \left( Y_N'^T {1 \over \Lambda'}Y'_N \right)_{\alpha\beta}, \,
c^{d=6}_{\alpha \beta} \equiv \left( Y_N^\dagger {1 \over \Lambda^\dagger \Lambda} Y_N \right)_{\alpha\beta}  + {\mathcal O}\left( {1 \over \Lambda'} \right).
\end{eqnarray}
Therefore, their flavour structures are completely unrelated and condition b) is not satisfied for these models.
Also, in contrast with type A models, the simplest case with $n=1$ does not lead here to a phenomenologically viable model since there is only one massive neutrino, and at least $n=2$ should have to be explored.

A possibility to enforce MFV in this case would be to  have both 
$\Lambda$ and $\Lambda'$ proportional to the identity matrix, and $Y_N 
\propto Y'_N$. This might be justified assuming for instance the flavour 
symmetry in eq.~(\ref{typeBsym2}). This  would not forbid however a 13 
entry in eq.~(\ref{MtypeB2}) proportional to  the $Y_N$ spurion, and 
additional small parameters would thus be required to ensure suppressed 
neutrino masses in this case. Note also that a spontaneously broken 
symmetry pattern  
cannot generate any 14 entry in eq.~(\ref{MtypeB2}) at the renormalizable 
level.

Finally, note that leptogenesis has been studied in some models of type B  in Ref.~\cite{leptosimple}, 
and in the ``extended MFV" framework in Ref.~\cite{isidoriporretti}.

\vspace{0.2cm}

In summary, the models of type B are interesting in particular as ultraviolet completions of MFV neutrino mass models of type A. They  involve two physical scales, associated to the masses of extra heavy fermions -SM singlets or triplets- and in them the approximate $U(1)_{LN}$ symmetry is recovered in the limit of large  $ \Lambda_{LN}$,  characteristic of some heavy fermion mass, and not by introducing very small mass terms or couplings. Although physically more appealing, the presence of two distinct mass scales is not stable under radiative corrections (unless some couplings are small),  which is nothing but the standard naturalness problem.

\section{Conclusions}

There are in the literature many minimal models which lead to predictions for the leptonic $c^{d=5}$ flavour structure,  assuming that some of the entries of the Yukawa coupling matrices and/or right-handed neutrino mass matrix vanish or are negligibly small \cite{zeros}. However, as these models typically  lead to very suppressed $c^{d=6}$ coefficients, they are not experimentally verifiable: in the seesaw model with three (two) right-handed neutrinos there is a nine(four)-dimensional space of parameters  which can lead to the same neutrino mass matrix. In order to be established, a model must lead to measurable effects other than neutrino masses and mixings. 
 The working hypothesis is that neutrino masses are generated by some new physics which decouples at low energies, leaving behind also a tower of $d\ge6$ effective operators. Particularly interesting and predictive models of this kind are the MFV models where not only the $d=6$ couplings are large, but can be determined from the neutrino mass matrix, up to an overall normalization scale.

We have  explored various realizations of the MFV hypothesis in the lepton sector \cite{mfv1}. We have argued that it  requires two a priori unrelated conditions. The first is the existence of some approximate $U(1)_{LN}$ lepton number symmetry implying two distinct scales, $\Lambda_{LN} \gg \Lambda_{FL}$: the first scale suppresses all operators violating lepton number, such as the $d=5$ Weinberg's operator, and the second one suppresses to a lesser extent flavour violating but lepton number conserving processes, such as $l_i\rightarrow l_j \gamma$, mediated by $d=6$ effective operators. The second requirement is the existence of a relation between  the flavoured coefficients of the corresponding effective operators. 

We did find explicit realizations of these hypotheses in the context of  seesaw scenarios. First, type II seesaw models (that is, scalar mediated) are of the type classified as {\it minimal} in Ref.\cite{mfv1}, in which the coefficient $c^{d=5}$ of  Weinberg's operator is the basic and only flavour spurion in the model.  The coefficients of the $d=6$ effective operators are then quadratic in this basic spurion. Second, we have considered seesaw models of type I -i.e. mediated by singlet or triplet fermions- with an approximate $U(1)_{LN}$ symmetry~\cite{u1}. Among them,  some of the so-called inverse seesaw models fall in the category of the {\it extended} models defined in Ref.~\cite{mfv1}, in which the $d=5$ and $d=6$ operators have identical flavour structure  in the absence of CP violation. 

The most interesting result of this work is that we have identified the simplest model (involving just two extra singlet or triplet fermions), which automatically satisfies the hypotheses of MFV. It is a seesaw model 
  in which the assumption of an approximate lepton number $U(1)_{LN}$ symmetry 
directly implies a relation between the flavour structures of the $d=5$ and the $d=6$ effective couplings.  The light neutrino mass matrix  involves in this model only one Majorana phase.
The flavour violating rates induced by the $d=6$ couplings can be reconstructed - including CP phases - from the parameters in the light neutrino mass matrix, except for: 1) a global normalization and 2) discrete degeneracies in the Majorana phase.
The relation between the $d=5$ and the $d=6$ operator coefficients in this case differs in nature from those previously considered~\cite{mfv1}.
 We presented the phenomenological implications of this simplest model  in what respects  the comparison of the $l_i \rightarrow l_j \gamma$ branching ratios and  neutrinoless double beta decay.  The model is 
a simple alternative for having sufficiently small neutrino masses, with large and fully predictable flavour violating effects. 

It remains to be seen whether it can be successful in explaining the origin of the matter-antimatter asymmetry. 

\section*{Acknowledgments}
We acknowledge inspiring discussions with C.~Biggio and F.~Bonnet. We thank A.~De R\'ujula 
for 
a critical reading of the first version of this paper. We also thank O.~Lychkovskiy and
M.~Vysotsky for
a
discussion
on the role of Majorana phases in the type-II model, and E.~Fern\'andez-Martinez and R.~Alonso for pointing out a discrete degeneracy.
M.B.~Gavela and D.~Hern\'andez 
 received partial support from CICYT through the project
FPA2006-05423, as well as from the Comunidad Aut\'onoma de Madrid
through Proyecto HEPHACOS; P-ESP-00346.  T.~Hambye thanks the FNRS-FRS and Belgian Science Policy (PAI VI/11) for support.  D.~Hern\'andez acknowledges financial support from the MEC through FPU grant AP20053603.
P.~Hern\'andez acknowledges partial finantial support from the research grant
FPA-2007-60323, the European projects EUROnu and FLAVIAnet, and the Consolider-Ingenio 2010 projects CUP and CPAN.

\newpage

\appendix


\section{Naturalness}


We address here the question of naturalness and the stability of the scales present in the models considered, which is an issue as they include at least one scale larger than the electroweak one.

In models of type A as in eqs.~(\ref{eq:newper}) and (\ref{MnuA}), the quantum corrections induced on the size of the electroweak scale by the presence of  $\Lambda$ of ${\mathcal{O}}(TeV)$  are not significant, because they have to be proportional to  the small parameters $\epsilon$ or $\mu, \mu'$. 

While the smallness of the $\epsilon$ entries in eq.~(\ref{eq:newper}) can be technically natural, as discussed in section~4, a naturalness problem arises instead in the type A models in eq.~(\ref{MnuA}), when the zeros are justified as due to a conserved global lepton number, which is then spontaneously broken, i.e. ~by the vev of a singlet scalar field $S$ with interactions given in eq.~(\ref{eq:typea_singlet}). The scalar potential $V(S,\phi)$, 
\be
V (S,\phi)= \lambda_\phi(\phi^\dagger \phi)^2 + \lambda_S(S^\dagger S)^2 + \mu_\phi^2\phi^\dagger\phi + \mu_S^2S^\dagger S + \lambda(\phi^\dagger \phi)(S^\dagger S)\,,
\ee
leads to
\be \label{vev-S}
<S> = \sqrt{ \frac{( \lambda \mu_\phi^2-2\lambda_\phi\mu_S^2)}{4\lambda_\phi\lambda_S - \lambda^2} }\,.
\ee
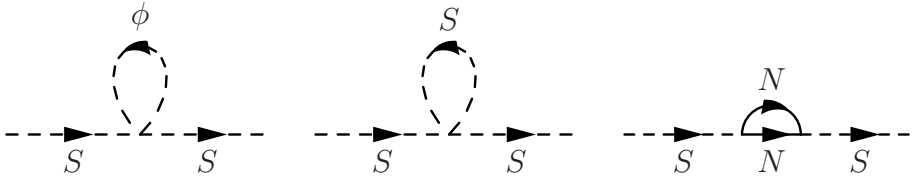
\begin{figure}[h] \label{feynm1}
\begin{center}
\begin{fmffile}{diag1a}
 \begin{fmfgraph*}(100,80)
   \fmfleft{i1}
   \fmfright{o1}
   \fmf{scalar,label= $S$}{i1,v1}
   \fmf{scalar,label=$\phi$}{v1,v1}
   \fmf{scalar,label=$S$}{v1,o1}
 \end{fmfgraph*}
\end{fmffile}
\begin{fmffile}{diag2a}
 \begin{fmfgraph*}(100,80)
   \fmfleft{i1}
   \fmfright{o1}
   \fmf{scalar,label= $S$}{i1,v1}
   \fmf{scalar,label=$S$}{v1,v1}
   \fmf{scalar,label=$S$}{v1,o1}
 \end{fmfgraph*}
\end{fmffile}
\begin{fmffile}{diag3a}
 \begin{fmfgraph*}(110,80)
   \fmfleft{i1}
   \fmfright{o1}
   \fmf{scalar,label= $S$}{i1,v1}
   \fmf{fermion,label=$N$}{v1,v2}
   \fmf{fermion,label=$N$,left=1}{v1,v2}
   \fmf{scalar,label=$S$}{v2,o1}
 \end{fmfgraph*}
\end{fmffile}
\caption{Loop corrections to mass of the scalar $S$ in type A models with spontaneously broken lepton number.}
\end{center}
\end{figure}

This vev has to be small compared to $\Lambda$, as $\mu=g <S>$, $\mu'=g'<S>$, see  eqs.~(\ref{MnuA}) and (\ref{eq:typea_singlet}). The problem arises  because, for instance, $\mu_S$ is destabilized at one-loop by contributions sensitive to high scales and only weighted by the couplings $g$, $g'$, $\lambda_S$ or $\lambda$. As an example, the contribution from the three diagrams in Fig.~6 are, respectively, 
\be
\delta \mu_S^2 \sim \frac{\lambda}{(4\pi)^2}\left[ \Lambda_{c}^2 - m_\phi^2\ln \left( 1 + \frac{\Lambda_c^2}{m_\phi^2} \right)\right]\,,
\ee
\be
\delta \mu_S^2 \sim \frac{3\lambda_S}{(4\pi)^2}\left[ \Lambda_c^2 - m_S^2\ln \left( 1 + \frac{\Lambda_c^2}{m_S^2} \right) \right]\,,
\ee
\be
\delta \mu_S^2 \sim \frac{(g+g')^2}{4(4\pi)^2} \left[ \Lambda_c^2 + \Lambda^2\ln\left( 1 + \frac{\Lambda_c}{\Lambda} \right) \right]\,,
\ee
where $\Lambda_c$ is a cutoff scale to be removed by renormalization, after which  finite contributions will still remain proportional to physical scales such as the Higgs mass $m_\phi$, the scalar mass $m_S$ or the flavour scale $\Lambda$. A fine-tuning is thus necessary to preserve the desired hierarchy, unless the dimensionless couplings $g$, $g'$, $\lambda_S$ and $\lambda$ turn out to be small.

Type B models involve at least two large scales, represented by $\Lambda$ and $\Lambda'$, typically with $\Lambda' \gg \Lambda$. The class of models in eq.~(\ref{mnutypeb}) taken by themselves is free from naturalness problems. To illustrate it, it suffices to take the simpler case $\mu_1=\mu_2=\Lambda$,
\begin{align}
\msc{L} & = \msc{L}_{SM} + i\bar{N}\slashed{\partial}N + i\bar{N'}\slashed{\partial}N' + i\bar{N''}\slashed{\partial}N'' - \Big[ Y_N\bar{N}\tilde{\phi}^\dagger {\ell}_L + \frac{\Lambda'}{2}\bar{N''}N^{''\,c} + \Big. \nn \\
& \quad\quad \Big. + \frac{\Lambda}{2}\Big( \bar{N}N^{''\,c} + \bar{N'}N^c + \bar{N}N^{''\,c} + \bar{N''}N^{c} + \bar{N''}N^{''\,c} + \bar{N''}N^{'\,c} \Big) +  \trm{h.c.} \Big]\,,
\end{align}
which becomes, in the basis of mass eigenstates denoted $N_1,\,N_2,\,N_3$,
\begin{align}
\msc{L} & = \msc{L}_{SM} + i\bar{N}_1\slashed{\partial}N_1 + i\bar{N}_2\slashed{\partial}N_2 + i\bar{N}_3\slashed{\partial}N_3 - \Big[ Y_N(\alpha \bar{N}_1 + \beta\bar{N}_2 + \gamma\bar{N}_3)\tilde{\phi}^\dagger L + \nn \Big. \\
& \quad\quad \Big. + \frac{\Lambda}{2} (\bar{N}_1N_1^c + \Lambda\bar{N}_2N_2^c) + \frac{\Lambda'}{2}\bar{N}_3N_3^c + \trm{h.c.} \Big]\,,
\end{align}
where $(N_1, N_2, N_3)^T = U\,(N, N',N'')^T$, $U$ being unitary. 
$\alpha$, $\beta$ and $\gamma$  are functions of $\Lambda$ and $\Lambda'$ which, up to order $\frac{\Lambda}{\Lambda'}$, read
\be
\alpha = \frac{i}{\sqrt{2}} \,, \quad \beta = -\frac{1}{\sqrt{2}}  \, ,\quad \gamma = \frac{\Lambda}{\Lambda'}\,.
\ee
In this basis, it is directly seen that the coupling of the Higgs to the heaviest field $N_3$ is suppressed by the factor $\frac{\Lambda}{\Lambda'}$, a fact that could already be guessed from eq.~(\ref{mnutypeb}). Also, for instance, the amplitude of the loop diagram depicted in Fig.~7, 
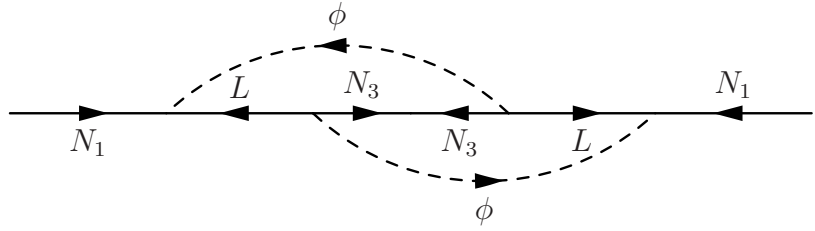
\begin{figure}\label{feynm2}
\begin{center}
\begin{fmffile}{diag3}
  \begin{fmfgraph*}(300,100)
    \fmfleft{i1}
    \fmfright{o1}
    \fmf{fermion,label=$N_1$,tension=5}{i1,v1}
    \fmf{fermion,label=$L$,tension=3}{v2,v1}
    \fmf{fermion,label=$N_3$}{v2,v3}
    \fmf{fermion,label=$N_3$}{v4,v3}
    \fmf{fermion,label=$L$,tension=3}{v4,v5}
    \fmf{fermion,label=$N_1$,tension=5}{o1,v5}
    \fmf{scalar,label=$\phi$,right=.4}{v4,v1}
    \fmf{scalar,label=$\phi$,right=.4}{v2,v5}
  \end{fmfgraph*}
\end{fmffile}
\end{center}
\caption{Two loop correction to the $N_1$ mass on type B model.}
\end{figure}
can be written as
\be \label{amplitude}
\mcl{M}(p)_A^C = i\frac{Y^4}{2}\frac{\Lambda^2}{\Lambda'}\int \frac{d^4l\,d^4k}{(2\pi)^8} \frac{l_\mu k_\nu \sigma^\mu_{A\dot{B}}(\bar{\sigma}^\nu)^{\dot{B}C}}{l^2k^2[(l+k-p)^2-\Lambda'^2](p-l)^2(p-k)^2},
\ee
where $p$ is the incoming momentum and where we have neglected the mass of the Higgs and the lepton running inside the loop. The integral in eq.\eqref{amplitude} yields a logarithmic contribution of order one, hence the suppression factor $\gamma^2= (\Lambda/\Lambda')^2$ guarantees no higher order correction to mass of the $N_1$. Furthermore, it is clear that this type of suppression always appears when the $N_3$ field runs inside a loop, and no naturalness problem results in this  model.

The trouble is that the zeros in eq.~(\ref{mnutypeb}) appear to be an ad hoc constraint. Again, they can be justified if lepton number is a symmetry of the Lagrangian, spontaneously broken by the vev of some scalar field(s), i.e. a singlet scalar $S$, to induce the entries $\mu_1$, $\mu_2$ while the null entries remain protected by the symmetry. This solution rises questions of naturalness, though, as quantum corrections may push the value of $\Lambda $ towards that of the higher scale $\Lambda'$.  We will illustrate it in what follows.

Let us promote the Lagrangian corresponding to eq.~(\ref{mnutypeb}) to the lepton number conserving one
\begin{align} 
\msc{L} & = \msc{L}_{SM} + i\bar{N}\slashed{\partial}N + i\bar{N'}\slashed{\partial}N'+ i\bar{N''}\slashed{\partial}N''\nn\\
&-V(S,\phi) - \left[ \frac{\Lambda}{2}(\bar{N'}N^c + \bar{N}N^{''\,c} + \frac{\Lambda'}{2}\bar{N''}N''^c + \right] \nn \\
& \quad\quad \left. + Y_N\bar{N}\tilde{\phi}^\dagger L +  \frac{f_1}{2}S(\bar{N'}N''^c + \bar{N''}N'^c) + \frac{f_2}{2}S^\dagger(\bar{N}N''^c + \bar{N''}N^c) + \trm{h.c.} \right]\,, 
\label{LwithScaseB}
\end{align}
where $S$ is a new scalar field with charge $-1$ under lepton number symmetry. Note that the symmetry is only violated after $S$ acquires a vev, resulting in $\mu_1$, $\mu_2$ in eq.~(\ref{mnutypeb}) given by  $\mu_1\equiv  f_1\,\langle S \rangle$ and $\mu_2\equiv  f_2\,\langle S\rangle$.  
Due to the couplings of the $S$ field new quantum corrections arise. The  diagram in Fig.~8 
\vspace{.2cm}
\begin{figure} \label{feynm3}
\begin{center}
\begin{fmffile}{diag2}
  \begin{fmfgraph*}(200,80)
    \fmfleft{i1}
    \fmfright{o1}
    \fmf{fermion,label=$N$}{i1,v1}
    \fmf{phantom}{i1,v1}
    \fmf{fermion,label=$N''$,label.side=left}{v2,v1}
    \fmfv{decoration.shape=cross}{v2}
    \fmf{fermion,label=$N''$}{v2,v3}
    \fmf{fermion,label=$N'$,label.side=left}{o1,v3}
    \fmf{scalar,label=$S$, left=1}{v1,v3}
    \fmf{phantom}{o1,v3}
    \end{fmfgraph*}
  \end{fmffile}
  \caption{Loop corrections to mass of $N_1$ in type B models with spontaneously broken lepton number.}
\end{center}
\end{figure}
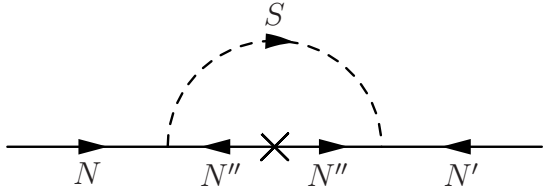
induces a  correction to the scale $ \Lambda$ given by
\be
\delta \Lambda \sim \frac{f_1f_2}{(4\pi)^2}\,\Lambda' .
\ee
where logarithms of order one have been neglected.
This correction could suffice to destabilize the $\Lambda$ scale. Note though that it does not need to be the case if the dimensionless coupling $f_2$, which does not enter  in eq.~(\ref{LtypeB}), turns out to be sufficiently small.

\vspace{.2cm}

In summary, naturalness issues arise in those models in which the justification of the vanishing or smallness of some couplings calls for a spontaneous breaking of lepton number symmetry. In the scenarios of this type analyzed, the problem can be evaded if certain dimensionless new couplings take small values. If this is the case, although we have not identified a symmetry reason justifying such small values, the protection of the size of the scales is technically natural.


\end{document}